\documentclass[superscriptaddress, a4paper,aps,twocolumn,nobibnotes,pra]{revtex4}
  \usepackage[utf8]{inputenc}
  \usepackage{svg}
  \usepackage{pgfplots} 
  \usepackage{ulem}
  \usepackage{color}
  \usepackage{graphicx}
  \usepackage{amssymb}
  \usepackage{amsmath}

  \usepackage{xcolor}
  \usepackage{bbold}
  \usepackage{wrapfig}
  \usepackage{bm}
  \usepackage{verbatim}
  \usepackage[utf8]{inputenc}
  \usepackage{bbm}
  \usepackage{transparent}

  \usepackage{gensymb}
  \usepackage{float}
  \usepackage{hyperref}
	
\newcommand{\ket}[1]{\mbox{$ | #1 \rangle $}}
\newcommand{\bra}[1]{\mbox{$ \langle #1 | $}}

\newcommand{\ee}{\end{eqnarray}}

\newcommand\preq{\mathrel{\overset{\makebox[0pt]{\mbox{\normalfont\tiny\sffamily Pr}}}{=}}}

\definecolor{ao(english)}{rgb}{0.0, 0.5, 0.0}
\definecolor{awesome}{rgb}{1.0, 0.13, 0.32}
\definecolor{amber}{rgb}{1.0, 0.75, 0.0}
\definecolor{babyblue}{rgb}{0.54, 0.81, 0.94}
\definecolor{babypink}{rgb}{0.96, 0.76, 0.76}
\definecolor{bluebell}{rgb}{0.64, 0.64, 0.82}
\pgfplotsset{compat=1.11,
        /pgfplots/ybar legend/.style={
        /pgfplots/legend image code/.code={
        \draw[##1,/tikz/.cd,bar width=3pt,yshift=-0.2em,bar shift=0pt]
                plot coordinates {(0cm,0.8em)};},
},
}

\begin{document}

\title{Equilibration of objective observables in a dynamical model of quantum measurements}

\author{Sophie Engineer}
\email[Email address: ]{sophiengineer@gmail.com}
\affiliation{Quantum Engineering Centre for Doctoral Training, H.\ H.\ Wills Physics Laboratory and Department of Electrical \& Electronic Engineering, University of Bristol, Bristol, United Kingdom}
\affiliation{Institute of Photonics and Quantum Sciences (IPAQS), Heriot-Watt University, Edinburgh, United Kingdom}

\author{Tom Rivlin}
\affiliation{Atominstitut, Technische Universit\"{a}t Wien, 1020 Vienna, Austria}

\author{Sabine Wollmann}
\affiliation{Institute of Photonics and Quantum Sciences (IPAQS), Heriot-Watt University, Edinburgh, United Kingdom}

\author{Mehul Malik}
\affiliation{Institute of Photonics and Quantum Sciences (IPAQS), Heriot-Watt University, Edinburgh, United Kingdom}

\author{Maximilian P. E. Lock}
\email[Email address: ]{maximilian.paul.lock@tuwien.ac.at}
\affiliation{Atominstitut, Technische Universit\"{a}t Wien, 1020 Vienna, Austria}
\affiliation{Institute for Quantum Optics and Quantum Information—IQOQI Vienna, Austrian Academy of Sciences, Boltzmanngasse 3, 1090 Vienna, Austria}
\medskip

\begin{abstract}
The challenge of understanding quantum measurement persists as a fundamental issue in modern physics. Particularly, the abrupt and energy-non-conserving collapse of the wave function appears to contradict classical thermodynamic laws. The contradiction can be resolved by considering measurement itself to be an entropy-increasing process, driven by the second law of thermodynamics. One such resolution explains the apparently irreversible emergence of objective outcomes in an isolated, unitarily-evolving quantum system via the theory of closed-system equilibration.
Working within this framework, we construct the set of \textit{`objectifying observables'} that best encode the measurement statistics of a system in an objective manner, and establish a measurement error bound to quantify the probability an observer will obtain an incorrect measurement outcome. Using this error bound, we show that the objectifying observables readily equilibrate on average under the set of Hamiltonians which preserve the outcome statistics on the measured system. Using a random matrix model for this set, we numerically determine the measurement error bound, finding that the error only approaches zero with increasing environment size when the environment is coarse-grained into so-called observer systems. This indicates the necessity of coarse graining an environment for the emergence of objective, classical measurement outcomes.
\end{abstract}

\maketitle

\section{Introduction}

The measurement of a quantum system is a key part of any experiment in quantum physics. While the process is well-modelled mathematically, the physical interpretation of this model stands as one of the most enduring challenges of modern physics~\cite{16brukner}. Additionally, the laws of thermodynamics, fundamental in our understanding of energy, entropy, and temperature, are seemingly violated~\cite{guryanova2020ideal} by the abrupt and energy-non-conserving collapse of the wave function~\cite{von2013mathematische, bassi2005energy,zurek2018quantum,carroll2021energy}. 

In this work we develop a model of the measurement process that offers a potential resolution to these issues, building on the work of~\cite{schwarzhans2025quantum}. The model hypothesises that quantum measurements emerge as a direct consequence of the second law, i.e.\ via an entropy-increasing transition. This would indicate that measurements are fundamentally thermodynamic processes. The idea can be combined with the well-established frameworks of decoherence and Quantum Darwinism~\cite{zurek2003decoherence,zurek2009quantum}, which investigate how some of the key features of a quantum measurement occur, to allow one to apply the tools of modern quantum statistical mechanics to model the irreversible emergence of classical measurement outcomes. 

In proposing this model we confront an important open question, sometimes called the ``big'' measurement problem~\cite{16brukner}: when should a process be modelled as unitary evolution, i.e.\ dynamically, and when should it be modelled as a measurement, i.e.\ instantaneous collapse? In the paradigm we describe here, the system being measured interacts with its surrounding environment, all processes are unitary, and it is the effect of the equilibration of observables, also known as equilibration on average~\cite{gogolin2016equilibration}, that generates seemingly irreversible measurements, just as classical statistical mechanics explains how seemingly irreversible phenomena arise from reversible dynamics.
A key result within this paradigm is that an exact, ideal projective measurement is impossible~\cite{schwarzhans2025quantum} (indeed, the infinite thermodynamic costs associated with them were already studied in detail in \cite{guryanova2020ideal}). However, as the environment grows in size, its encoding of the measurement statistics of the system (i.e.\ the ensemble of environmental states conditioned on the system pointer states) asymptotically approaches complete distinguishability, and hence the measurements can be treated as ideal in the sense of \cite{guryanova2020ideal}. Further, by coarse graining environmental subsystems into ``observer systems'', the approach to complete distinguishability is exponential with respect to the number of constituents of the observer system.

In~\cite{schwarzhans2025quantum}, the structure of the equilibrium state was investigated, specifically the extent to which a system's environment unambiguously encodes the measurement outcome. There it was shown that equilibration can lead to an approximate form of a state structure called Spectrum Broadcast Structure (SBS), as described in detail in Section~\ref{secObjectivity}. SBS implies that certain criteria of classical objectivity hold~\cite{korbicz2014objectivity, horodecki2015quantum}, as well as a strong form of Quantum Darwinism~\cite{le2019strong}. It has been well-studied in a number of examples~\cite{tuziemski2015objectivisation, tuziemski2015dynamical,tuziemski2016analytical,mironowicz2018system,kicinski2021decoherence,lee2024encoding}. It remains, however, to be understood which environment observables actually reveal these outcomes, and whether these observables equilibrate, which must be addressed if this paradigm can lead to a consistent model of measurement.

Naturally, the idea that quantum measurements are linked to irreversibility and an increase in entropy has been discussed before. But despite it having been studied by many well-known names in the history of quantum theory~\cite{64Szilard, 80Peres, 83Zurek,90Bell}, the role of this entropy increase (i.e.~tendency towards equilibrium) in driving the measurement process remains unclear. Even a well-known recent review of the measurement problem~\cite{16brukner} does not discuss entropy or statistical mechanics in relation the problem of irreversibility. Progress has however been made in quantify the thermodynamic costs of quantum measurements~\cite{busch1996quantum, 13Hormoz, 13AllahverdyanBalianNieuwenhuizen, 17ElouardHerreraAuffeves, guryanova2020ideal,25WallsBlossFord}, studying the feasibility of extracting information from decohering environments \cite{22TouilYanGirolami}, and the entropy production in objective collapse models~\cite{23ArtiniPaternostro,25ArtiniLoMonacoPaternostro}. Here, on the other hand, we adopt the perspective that it is in fact the tendency towards equilibrium itself that is responsible for the irreversibility of quantum measurements~\cite{schwarzhans2025quantum} and the quantum-to-classical transition. Given this, we study the extent to which this is compatible with the theory of the equilibration of isolated systems.

In particular, adopting objectivity~\cite{korbicz2021roads} as a marker of classicality, we construct a set of observables that best recover the extent to which the statistics of the measured system (in the measurement basis) is encoded into its environment, calling them the \textit{objectifying observables}. In order to quantify how effective these observables are at extracting measurement outcomes at equilibrium, we then construct a measurement error bound. This error quantifies the probability that an observer system will not accurately reproduce the measurement outcome. It can be concluded that any dynamics leading to a vanishing error bound for multiple observers are measurements, in the sense that they satisfy the requirements of objectivity, stability and irreversibility. Here we also require that the dynamics only minimally disturbs the system itself in its pointer basis.   

Both the objectifying observables and the error bound that we construct are independent of the form of the system-environment dynamics. The equilibration of an observable, however, depends on the dynamics, and we investigate this for the structure of system-environment interactions considered in \cite{korbicz2017generic,mironowicz2017monitoring,schwarzhans2025quantum}. 
We then use numerical simulations of a random matrix model \cite{mehta2004random} to investigate the conditions under which the measurement error is minimal. In particular, we consider chaotic conditional interactions modelled by a Gaussian random matrix ensemble~\cite{DAlessio2016}.

We find that the observables which most readily encode the measurement statistics of the system also readily equilibrate for large environment dimensions. Importantly, we also find that for the chaotic conditional interactions that we study, the distinguishability of measurement outcomes can only be approached via the coarse graining of the environment into large observer systems. 

We begin by reviewing the notion of objectivity (in a strict sense defined in the following section) and the equilibration of isolated quantum systems, i.e.\ equilibration on average -- it is the combination of these two concepts that drives this work. We then define the set of observables that most readily encode the statistics of the system and derive an error bound on an observer's ability to extract measurement information through an equilibration process. Following this, we investigate the size of the error bound, averaging over random dynamics described by a broadcasting Hamiltonian. Finally, we consider different initial states and the impact they have on the error, and we discuss the ramifications of the results.

\section{Objectivity via dynamical equilibration of closed systems}

\subsection{Objectivity of measurement outcomes} \label{secObjectivity}

\begin{figure}[t!]
\centering\includegraphics[width=0.5\textwidth]{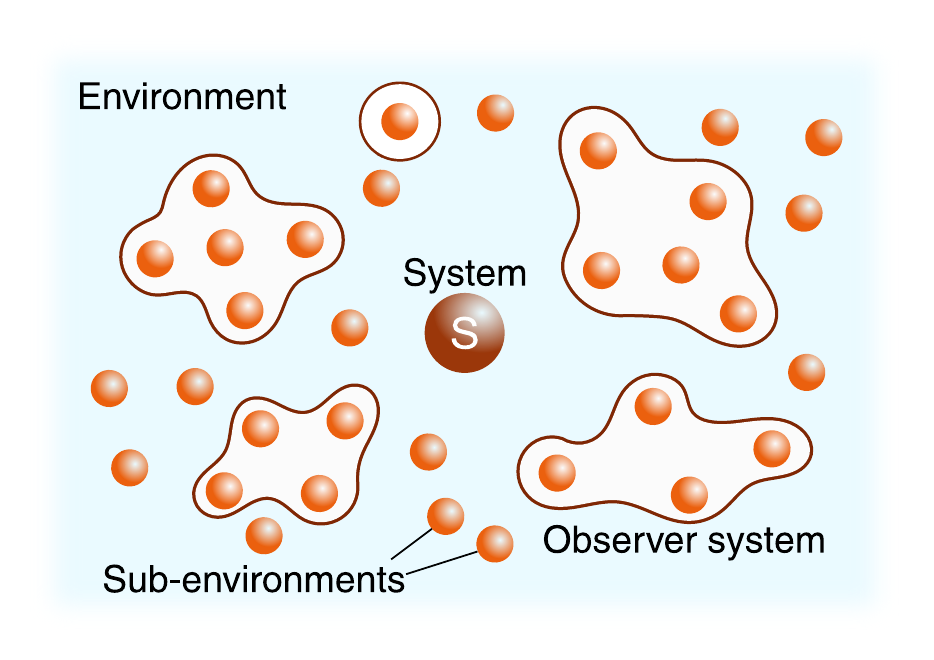}
\caption{Conceptual illustration. We model a measurement as an isolated system-environment interaction. The system being measured, $S$, interacts unitarily with an environment (blue background). The environment is composed of many sub-environments (orange spheres), and observer systems are collections of sub-environments (white envelopes). We imagine a scenario in which each observer has access to an observer system.}
\label{fig:schematic_conceptual}
\end{figure}

Any reasonable model of a quantum measurement must reproduce its key features: irreversibility, and the stability and objectivity of outcomes. Objectivity of measurement outcomes was defined in the context of Quantum Darwinism (QD) as information that is independently accessible to multiple observers, without perturbing the system under consideration \cite{zurek2009quantum}. Classical mechanics satisfies this definition, whereas in quantum mechanics, properties of a system are not objective until after a measurement has occurred. A model of quantum measurement must therefore be able to explain how non-objective quantum states can irreversibly produce objective measurement outcomes. In the conventional picture of quantum measurement as a `wavefunction collapse', post-measurement objectivity is imposed as a postulate. The theory of QD was developed to understand how our objective classical world emerges from non-objective quantum systems, without imposing this assumption `from above', and has been investigated in many physical scenarios, both theoretically \cite{blume2008quantum,riedel2010quantum,riedel2011redundant} and experimentally \cite{unden2019revealing,kwiatkowski2021appearance, ciampini2018experimental,chen2019emergence,chisholm2021witnessing}. In this paradigm, the coupling between the system and the observer systems is uncontrolled and time-independent. This contrasts with, models such as the von Neumann model, which considers a ``premeasurement''~\cite{busch1996quantum}, namely a coherently-controlled interaction over a finite length of time that couples the system of interest to a single degree of freedom of a detector. The observer systems described above subsumes such a detector -- the measurement apparatus is any complex system coupled to the system of interest in a way that is conditional in the measurement basis.

Following the QD formalism, the notion of objectivity was formalised into a state structure, namely Spectrum Broadcast Structures (SBS) \cite{korbicz2014objectivity, horodecki2015quantum}, defined in Eq.~\eqref{eSBSdefn} below. This state structure is equivalent to so-called `strong Quantum Darwinism', combined with `strong independence' of environmental subsystems (meaning there are no correlations between them, except for shared information about the system) \cite{le2019strong}. Following the introduction of SBS states, there has been much research into their appearance in known physical models such as Brownian motion \cite{tuziemski2016analytical, tuziemski2015objectivisation, tuziemski2015dynamical}, spin-spin models~\cite{kicinski2021decoherence, mironowicz2018system} and a boson-spin model \cite{lee2024encoding}. Beyond these physical models, both dynamics with self-evolution \cite{mironowicz2022non} and the relationship between entanglement and objectivity \cite{roszak2019entanglement} have been studied, as well as the connection between objectivity and thermodynamic constraints \cite{debarba2024broadcasting,25CandeloroDebarbaBinder}. The generic emergence of objectivity was investigated by averaging over dynamics to see typical behaviours \cite{korbicz2017generic} and recent work even proposed a Bell-like inequality to witness non-objectivity \cite{poderini2022witnessing}. 

As illustrated in Fig.~\ref{fig:schematic_conceptual}, we consider some system of interest $S$ interacting with an `environment', where this environment consists of $N_{E}$ quantum systems, which we call `sub-environments'. We collect these sub-environments into $N_{O}$ groups, which we call `observer systems', and for simplicity, we will assume that each observer system contains $n$ sub-environments (though our results can be readily generalised to the case where each observer system contains different numbers of sub-environments). For example, the environment could include a physical measurement device such as a single-photon detector, along with its surroundings in a laboratory. The observer systems would then be distinct fractions of this.

Let us denote the initial state of $S$ by $\rho_{S,0}$ and the measurement basis by $\lbrace\ket{i}_{S} \rbrace_{i=1,2\dots d_{S}}$, with $d_{S}$ denoting the dimensionality of the Hilbert space of the system $S$ (note that we have assumed a non-degenerate observable for simplicity). The probability associated with the measurement outcome $i$ is then $p_{i}= \bra{i} \rho_{S,0}\ket{i}_{S}$. Demanding that, at some later time, the observer systems objectively (as defined in~\cite{korbicz2014objectivity}) encode the state of $S$ in the measurement basis implies that their combined state at that time has spectrum broadcast structure (SBS): 
\begin{align} \label{eSBSdefn}
    \rho^{\mathrm{SBS}} =\sum_{i=1}^{d_{S}} p_{i} \ket{i}\! \bra{i}_{S} \bigotimes_{k=1}^{N_{O}}\rho_k^{(i)},
\end{align}
where for all observer systems $k$, the conditional states are completely distinguishable: $\rho_k^{(i)}\rho_k^{(j)}=0$, $\forall$ $i\neq j$~\cite{korbicz2021roads}. This is equivalent to $F\left(\rho_k^{(i)}, \rho_k^{(j)}\right) = 0$, where $F(\rho, \sigma)=\left(\operatorname{tr} \sqrt{\sqrt{\rho} \sigma \sqrt{\rho}}\right)^2$ is the fidelity, which in turn implies the existence of a protocol which can perfectly distinguish states $\rho_k^{(i)}$ and $\rho_k^{(j)}$ (see e.g.~\cite{kargin2005chernoff}). 

In this paradigm, observers are each associated with an observer system, and the distinguishability of the conditional states of the $k^\text{th}$  observer system $\lbrace\rho_{k}^{(i)}\rbrace_{i=1,2\dots d_{S}}$ corresponds to the $k^\text{th}$ observer's ability to distinguish, correctly and with certainty, between the possible system states $\{\ket{i}_{S}\}_{i=1,2\dots d_{S}}$.

A question that naturally follows this idea of objectivity is, given a general multipartite state, with a defined system that has been measured, how far away is it from being an objective SBS state? There are many ways to answer this question: one could use the trace distance as a distance metric between states, as in \cite{mironowicz2017monitoring, mironowicz2022non}. However, as the set of SBS states is not convex, minimising the trace distance to the closest SBS state is not efficient. An alternative idea, discussed and used in \cite{mironowicz2017monitoring, le2018objectivity, poderini2022witnessing}, is to calculate a probability of success, i.e.\ asking what the probability is that an observer, attempting to extract the measurement outcome from an observer system, would obtain the correct result. This is essentially a state discrimination problem \cite{montanaro2008lower} and we outline this approach in Section~\ref{measure}. 

\subsection{Objectivity through equilibration} 

We investigate a model of measurements where the emergence of an objective post-measurement state occurs spontaneously through unitary dynamics, as the result of closed-system equilibration~\cite{gogolin2016equilibration}. In this notion of equilibration, one does not demand that a state tends inexorably to an equilibrium state, but rather that the statistics of certain observables or subsystems are on average represented by the equilibrium state, which can be calculated via the infinite-time average~\cite{short2011equilibration}:
\begin{equation} \label{eEquilibState}
    \rho_{\text{eq}} = \lim_{T\rightarrow\infty} \frac{1}{T}\int_{0}^{T} \rho(t) dt = \sum_{n} P_{n}\rho(0)P_{n} ,
\end{equation}
where $P_{n}$ is the projector onto the eigenspace corresponding to the $n^\mathrm{th}$ eigenvalue of the Hamiltonian, and $\rho(t) = U(t)\rho(0)U^{\dagger}(t)$ is the time-evolution of the initial system-plus-environment state $\rho(0)$. The equilibrium state maximises the von Neumann entropy given the constants of motion~\cite{gogolin2016equilibration,romero2020equilibration}. Specifically, we say that an observable \textit{equilibrates on average} when, for `most' times $t$, a measurement of the observable will follow statistics close to those of the equilibrium state $\rho_{\text{eq}}$~\cite{reimann2008foundation}. Hence, the state $\rho(t)$ becomes very `close' to the equilibrium state $\rho_{\text{eq}}$ with respect to equilibrating observables. These observables tend to be highly non-commuting with the Hamiltonian~\cite{anza2018degenerate}, thus they are not strictly constants of motion.

Bounds on the proximity of the actual state to the equilibrium state, from the perspective of the observable in question, are well-known \cite{reimann2008foundation,short2011equilibration,short2012quantum}; one such bound states that for an arbitrary observable $O$ and a Hamiltonian with non-degenerate energy gaps~\cite{reimann2008foundation,short2011equilibration,short2012quantum},  
\begin{align}
    \left\langle|\operatorname{tr}[\rho(t) O]-\operatorname{tr}[\rho_{eq} O]|^2\right\rangle_{\infty} \leqslant \frac{\|O\|^2}{d_{\mathrm{eff}}},
\end{align}
where we use the notation $\langle \cdot \rangle_{\infty} = \lim_{T\rightarrow \infty} \int_{0}^{T} \cdot\ dt/T$ to denote the infinite-time average, $\|\cdot\|$ is the standard operator norm (the largest singular value of the operator) and $d_{\mathrm{eff}}$ is the effective dimension (sometimes called the inverse participation ratio). The effective dimension depends on the probability for each energy eigenstate of the Hamiltonian $H = \sum \lambda_{n} P_{n}$ to be occupied by the initial state \cite{romero2020equilibration}, and is defined as:
\begin{equation} \label{eEffDim}
    d_{\mathrm{eff}}^{-1}(\rho):=\sum_{n=1}^{\mathfrak{D}}\left(\operatorname{tr}\left[P_n \rho(0)\right]\right)^2,
\end{equation}
where $\mathfrak{D}$ is the dimensionality of the total Hilbert space. When the Hamiltonian is non-degenerate or the initial state is pure, the effective dimension is simply the inverse purity of the equilibrium state $d_{\mathrm{eff}}=1/\operatorname{tr}(\rho_{\text{eq}}^2)$ \cite{romero2020equilibration}. Heuristically, it tells us how much of the total Hilbert space is explored by the state during the time evolution. 

This bound indicates that for $\|O\|^2<<d_{\mathrm{eff}}$, the observable will equilibrate. A similar bound also exists for the equilibration on average of sub-systems \cite{linden2009quantum}, which can be proven using the observable bound. Note that the assumption of non-degenerate energy gaps can be relaxed~\cite{short2012quantum}.

As noted above, we model a quantum measurement as a dynamical and entropy-increasing transition to equilibrium. In \cite{schwarzhans2025quantum}, the conditions under which an equilibrium state is objective were studied, i.e.\ when $\rho_{\text{eq}}=\rho^{\text{SBS}}$. It was shown that this exact relation is impossible, and can only be approached asymptotically as the size of the environment increases. This leaves open the question, however, of which degrees of freedom (i.e. which observables) in the environment will encode the measurement statistics. We address this in Section~\ref{measure}. 

\section{Measurement via objectifying observables} \label{measure}

Let us now consider some arbitrary equilibrium state $\rho_{eq}$, calculated according to Eq.~\eqref{eEquilibState} for some time-independent Hamiltonian. Fixing a pointer (i.e.\ measurement) basis on the system $\{\ket{i}_{S}\}_{i}$, one can separate this state into diagonal and off-diagonal parts:
\begin{align} \label{eEquStateSeparateForm}
    \rho_{eq} =\sum_{i=1}^{d_{S}} q_{i} \ket{i}\! \bra{i}_{S} \otimes \rho^{(i)} + \gamma_{SE},
\end{align}
where the state $\rho^{(i)}$ is a density matrix on $\bigotimes_{k=1}^{N_{O}} \mathcal{H}_{k}$, which is the environment Hilbert space consisting of $N_{O}$ observer systems. Our goal is to quantify how well observers can distinguish between system states $\{\ket{i}_{S}\}_{i}$. To do this, we can use convex optimisation to find the optimal projectors for each observer system to distinguish between the possible outcomes of the measured system $S$ \cite{le2018objectivity,mironowicz2017monitoring}. In particular, we can define a set of projectors $\{\Pi_{k}^{i}\}_{i}$ via an optimisation procedure. By considering the maximum probability of success for each observer system $k$ to reproduce the statistics of $S$ (in the pointer basis), we can define a quantity $\mathcal{P}_{k}$ as \cite{poderini2022witnessing}:
\begin{equation} \label{prob_succ}
    \mathcal{P}_{k} = \max_{\{\Pi_{k}^{i}\}_{i}} \sum_{i=1}^{d_{S}} q_{i} \operatorname{tr}\left(\rho_{k}^{(i)}\Pi_{k}^{i}\right), 
\end{equation}
where $\rho_{k}^{(i)} = \operatorname{tr}_{k'\neq k}\left( \rho^{(i)} \right)$ is the state of the $k^{\text{th}}$ observer system conditioned on the system $S$ being in the state $\ket{i}$, and $\{\Pi_{k}^{i}\}_{i}$ are a complete set of projectors on $\mathcal{H}_{k}$. In the following, we will use $\{\Pi_{k}^{i}\}_{i}$ to denote the particular set of optimal projectors satisfying the maximisation in Eq.~\eqref{prob_succ}. We will use these optimal projectors to construct the objectifying observables, as well as an SBS state close to the given equilibrium state $\rho_{eq}$.

The probability of error for each observer attempting to ascertain the outcome of the measurement of $S$ will be quantified. Given a fictional scenario in which a system-environment equilibrates to an equilibrium state that is a perfect SBS state, the error is zero for all $k$ because one can always find optimal projectors that perfectly distinguish between the conditional states of each observer system. This means that all observers can correctly reproduce the statistics of the measured system and moreover that they all agree with each other. 

In reality, there are two sources of error to consider. For the first part, we know that equilibration is not an exact process. The statistics of an equilibrating observable may be close to those described by $\rho_{\text{eq}}$ for most times but there is still a finite difference between them at any given $t$, and for certain $t$ this difference may even be large (e.g. fluctuations from equilibrium). So even if the equilibrium state is a perfect SBS state, one has to account for this difference. This is addressed using time-averaged equilibrium bounds \cite{short2011equilibration,linden2009quantum}. Secondly, we know from \cite{schwarzhans2025quantum} that the equilibrium state cannot be an exact SBS state and for large distances between the equilibrium state and an SBS state, the probability for each observer to correctly encode the statistics of the system will be low. 

Operationally speaking, we only need to consider how well an observable can distinguish between the time-evolving state $\rho(t)$ and an SBS state that is close to the equilibrium state, which we denote $\rho_{\text{eq}}^{\text{SBS}}$. In Appendix~\ref{SM:candidate_SBS}, we show how to construct such an equilibrium-proximate SBS state. The observable equilibration bound from \cite{short2011equilibration} then provides a way to quantify the operational error arising from the difference between $\rho(t)$ and $\rho_{\text{eq}}^{\text{SBS}}$. 

The optimal projectors $\{\Pi_{k}^{i}\}_{i}$ defined via Eq.~\eqref{prob_succ} then allow us to define an observable for each observer system $k$. These observables optimally distinguish between the states of $S$. In particular, for each observer system $k$, we define the \textit{objectifying observable} $\hat{O}_{k}$:
\begin{equation} \label{obs}
    \hat{O}_{k} = \sum_{i=1}^{d_S} c_{i,k} O_{k}^{i} = \sum_{i=1}^{d_S} c_{i,k} \mathbb{1}_{S} \otimes \Pi_{k}^{i} \otimes \mathbb{1}_{k' \neq k}.
\end{equation}
Here, $O_{k}^{i}=\mathbb{1}_{S} \otimes \Pi_{k}^{i} \otimes \mathbb{1}_{k' \neq k}$ is an operator defined such that the optimal projector $\Pi_{k}^{i}$ is applied to the $k^{\text{th}}$ observer system, and the identity is applied to to all other observer systems and the measured system $S$. These projectors $O_{k}^{i}$ are weighted by some values $c_{i,k}$, which would correspond to observable outcomes when observing an actual physical system.   

To quantify the extent to which the objectifying observables encode measurement outcomes in an objective manner, we make use of a measurement-dependent distinguishability of states. Specifically, the distinguishability between two states, $\rho_{1}$ and $\rho_{2}$, when one has access to a specific measurement $M$, is defined as \cite{short2011equilibration}:
\begin{equation}
    D_M\left(\rho_1, \rho_2\right) \equiv \frac{1}{2} \sum_i\left|\operatorname{tr}\left(M_i \rho_1\right)-\operatorname{tr}\left(M_i \rho_2\right)\right|.
\end{equation}
Here, $M$ is a Positive Operator-Valued Measure (POVM), which is determined by a set of positive operators $\lbrace M_{i} \rbrace$, one for each outcome $i$, such that $\sum_{i} M_{i} = \mathbb{1}$. The distinguishability $D_M\left(\rho_1, \rho_2\right)$ then determines the probability of success in distinguishing $\rho_1$ from $\rho_2$ using the measurement $M$~\cite{short2011equilibration}. Note that this definition of distinguishability depends only on the $\lbrace M_{i} \rbrace$, and not on any observed values associated with each outcome. This means that the observable values $c_{i,k}$ in Eq.~\eqref{obs} play no role in the derivation of the following error bound.

We assume that each observer has access to the objectifying observable $\hat{O}_{k}$ defined in Eq.~\eqref{obs}. For each observer system $k$, we can look at the distinguishability between $\rho(t)$ and $\rho_{\text{eq}}^{\text{SBS}}$ according to $\hat{O}_{k}$. Let $\mathcal{E}_{k}$ denote the time-average of this distinguishability over the course of the evolution. This quantity, which we call the measurement error, satisfies the following bound
\begin{align}  \label{error}
    \mathcal{E}_{k} &= \left\langle D_{\hat{O}_{k}}\left(\rho(t), \rho_{\text{eq}}^{\text{SBS}} \right)  \right\rangle_{\infty} \notag\\
    &\leq  \mathcal{E}_{\text{obj}} +  \mathcal{E}_{\text{eq}},
\end{align}
where the two terms correspond to a lack of objectivity of the equilibrium state and a lack of equilibration respectively:
\begin{align} \label{eq:F_term}
    \mathcal{E}_{\text{obj}}  &= \sum_{i \neq j} \sqrt{q_i q_j}   F\left(\rho_k^{(i)}, \rho_k^{(j)}\right) + \Delta P ,  \\ \label{eq:D_term}
 \mathcal{E}_{\text{eq}}  &= 
 \frac{d_{S}}{4 \sqrt{d_{\mathrm{eff}}}}.
\end{align}
The terms $\{q_i\}_{i}$ and $\rho_{k}^{(i)} = \operatorname{tr}_{k'\neq k}\left( \rho^{(i)} \right)$ are defined from the general equilibrium state in Eq.~\eqref{eEquStateSeparateForm} and we recall that $\rho_{\text{eq}}^{\text{SBS}}$ is an SBS state close to $\rho_{\text{eq}}$, defined in Appendix~\ref{SM:candidate_SBS}. The term $\Delta P := P_{\{E_{k}^i\}}^{\text{Succ}} - P_{\{\Pi_{k}^i\}}^{\text{Succ}}$ is the difference between the probability of success when optimising over all POVM elements $\{E_{k}^{i}\}$ versus optimising over projectors (PVM elements) $\{\Pi_{k}^{i}\}$. We point out that in the case when the fidelity between two states is zero, the optimal measurement is a projective measurement~\cite{barnett2009quantum}. Therefore if $F\left(\rho_k^{(i)}, \rho_k^{(j)}\right) = 0$ for all $i \neq j$, then necessarily $\Delta P = 0$. Additionally, in the case when there are only two states in such a state discrimination problem, i.e.\ when $d_S=2$, the optimal POVM is again a projective measurement. As such, $\Delta P$ is identically zero in the numerical cases we will consider below. The effective dimension (defined in Eq.~\eqref{eEffDim}) satisfies $1\leq d_{\mathrm{eff}} \leq d_{S} d_{E}$ where $d_{S}$ denotes the dimension of the system being measured and $d_{E}$ the total environment dimension. See Appendix~\ref{SM_error_derivation} for the proof of this bound and more details on the term $\Delta P$. 

The inequality in~\eqref{error} quantifies the extent to which an arbitrary evolution can be understood as a measurement in the following sense. We are given some initial state, its unitary evolution and the choice of measurement basis, which combine to uniquely determine the decomposition in Eq.~\eqref{eEquStateSeparateForm}. Then, for some partitioning of the environment,
the inequality bounds from above how close (on average) the evolving state of the system is to the equilibrium-proximate SBS state $\rho_{\text{eq}}^{\text{SBS}}$, with respect to a set of optimal observables on that partition. This ``best-case scenario'' can then be understood as an upper bound on how well the measurement information has been objectively encoded into the environment due to the process of equilibration, or in short, whether the situation corresponds to a measurement (in the chosen basis) at equilibrium. As the inequality in~\eqref{error} shows, this bound is the sum of two effects: a) the failure of the equilibrium state to distinguish between different measurement outcomes, and b) the failure of the state to equilibrate on average in the first place.

To phrase it a different way, Eq.~\eqref{error} demonstrates the conditions under which the objectifying observables equilibrate and unambiguously encode the state of the system (in the measurement basis). First, the smaller the term $\mathcal{E}_{\text{obj}}$ is, the more the environment itself is able to encode the information at equilibrium, with $F(\rho_k^{(i)}, \rho_k^{(j)})\approx 0$ $\forall i,j,k$ with $i \neq j$ and $\Delta P \approx 0$ implying that the information is encoded unequivocally. (These assumptions should hold for reasonable models of measurement where the number of outcomes is far smaller than the total Hilbert space dimension.) Second, the system $S$ must equilibrate on average under the dynamics, i.e.\ ${\frac{d_{S}}{4 \sqrt{d_{\mathrm{eff}}}}\approx 0}$ (see Eq.~(16) in~\cite{short2011equilibration}), with $\mathcal{E}_{\text{eq}}$ quantifying how far from equilibrium the system is.

Note, however, that even when the average error $\mathcal{E}_{\text{obj}} +  \mathcal{E}_{\text{eq}}$ is negligible, the objectifying observables only accurately encode the measurement statistics of the initial state $\lbrace p_i =\langle i |\rho_{S,0}| i\rangle_{S} \rbrace_{i=1,2\dots d_{S}}$ when the equilibrium state itself encodes those statistics, i.e.\ when $q_i\approx p_i$ $\forall i$. Exact equality, $q_i=p_i$ $\forall i$, implies the Hamiltonian commutes with the observable being measured on the system of interest, i.e.\ that $H=\sum_{i=1}^{d_{S}} \ket{i}\!\bra{i}_{S}\otimes H^{(i)}_{E}$ for some $\lbrace H^{(i)}_{E}\rbrace_{i=1,2\dots d_{S}}$.

In the introduction we stated as a hypothesis that quantum measurements are driven by the tendency of systems to increase in entropy as they seek equilibrium. Encoding this as a process bringing the system-environment complex to an equilibrium state satisfying the property of objectivity, the inequality in~\eqref{error} quantifies the extent to which this hypothesis is satisfied in a given scenario. Thus, for a thermodynamically-driven, purely unitary measurement, the right hand side of the inequality in~\eqref{error} should be small -- the observables that optimally extract measurement information about the system from the environment should equilibrate well in the equilibration-on-average sense. Thus we can test the hypothesis by considering the behaviour of this upper bound in a scenario which we expect to behave like a measurement, as we do in the following section.

\section{Equilibration of objectifying observables under a broadcasting Hamiltonian} \label{SecGeneralBroadcasting}

In the previous section, we defined the measurement error as the lack of objectivity associated with an objectifying observable, averaged over the entire dynamics, and we derived an upper bound to this error. Here we wish to test the hypothesis that as the observer system dimension $d_k$ increases, we can expect, in general, for our error bound to decrease. This would then suggest that in the thermodynamic limit of large environment size (and hence the classical limit), the probability of error (or `non-objectivity') would be negligible. This naturally motivates a study of the measurement error, averaged over random interaction Hamiltonians. In the spirit of this, we consider the broadcasting Hamiltonian investigated in \cite{korbicz2017generic,mironowicz2017monitoring,schwarzhans2025quantum}, to wit:
\begin{equation} \label{ham}
    H=\sum_{i=1}^{d_S} \ket{i}\!\bra{i}_{S} \otimes \sum_{l=1}^{N_{E}}  H_l^{(i)},
\end{equation}
where each conditional Hamiltonian $H_{l}^{(i)}$ is drawn from the Gaussian Unitary Ensemble (GUE) \cite{dyson1962statistical,mehta2004random} and acts on a single sub-environment, indexed by $l$. We use this random matrix model to represent a generic, chaotic evolution of the observer system (see e.g.~\cite{DAlessio2016} for an extensive discussion of quantum chaos and random matrix models), thus limiting the extent to which specific features of chosen conditional Hamiltonians restrict our analysis, and allowing us to capture generic features of the dynamics. The form of the Hamiltonian in Eq.~\eqref{ham} follows from the requirements that the interaction preserves the pointer basis (see Section~\ref{measure}), i.e. that $q_i=p_i$ $\forall i$, and from the strong independence condition~\cite{horodecki2015quantum}.

\begin{figure*}[t!]
\centering\includegraphics[width=1\textwidth]{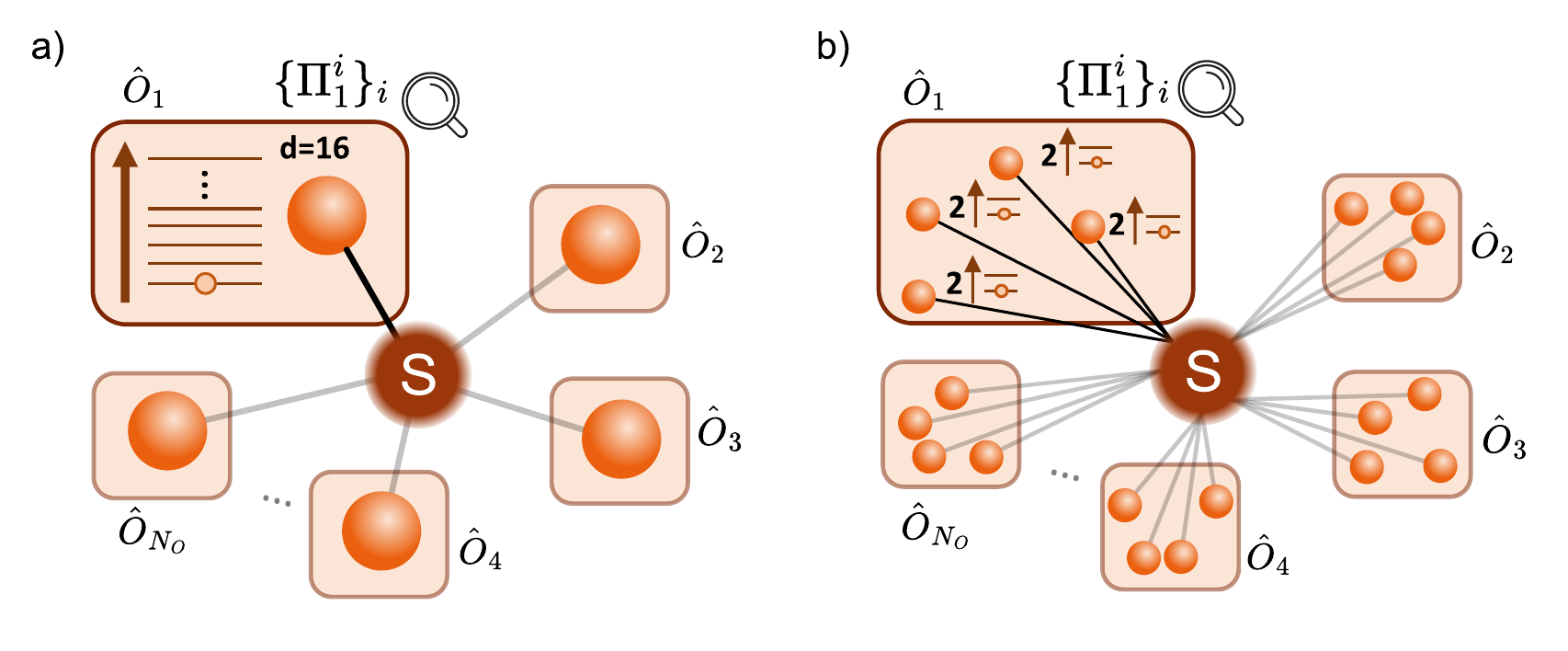}
\caption{Schematic illustrating two system-environment configurations. In both cases, we model a system $S$ being measured via a unitary interaction with its surrounding environment. The environment is composed of many sub-environments (orange spheres). The observer systems (indicated by shaded boxes) are collections of one or more sub-environments. Objectifying observables, corresponding to $\lbrace\Pi_{k}^{i}\rbrace_{i}$ are associated with each observer system. The black lines connecting the system to the sub-environments indicate interactions via the broadcasting Hamiltonian (Eq.~\eqref{ham}). In (a), we consider observer systems of single high-dimensional qudit sub-environments, where each qudit has dimension $d$ (here $d=16$). In (b) we consider observer systems of $n$ qubit sub-environments ($n=4$ in this illustration and so the total dimension of each observer system is $2^4 = 16$).   }
\label{fig:schematic}
\end{figure*}

Level repulsion in Gaussian random matrix models ensures that there is vanishing probability for eigenvalues to coincide, and we can therefore take each conditional Hamiltonian to be non-degenerate. Thus we can write each one as $H_{l}^{(i)} = \sum_{n=1}^{d_{l}} E_{n_{l}}^{(i)} \ket{E_{n_{l}}^{(i)}}  \bra{E_{n_{l}}^{(i)}}$, where $\ket{E_{n_{l}}^{(i)}}$ is the eigenstate corresponding to the unique eigenvalue $E_{n_{l}}^{(i)}$ and $d_{l}$ denotes the dimensionality of the sub-environment $l$.

In \cite{schwarzhans2025quantum}, it was shown that for an arbitrary, uncorrelated initial state $
\rho(0)=\rho_{S,0} \otimes_{l=1}^{N_{E}} \tilde{\rho}_{l,0}$, the equilibrium state, according to evolution via Eq.~\eqref{ham}, will be of the form 
\begin{equation} \label{rho_inf}
\rho_{\text{eq}}=\sum_{i=1}^{d_{S}} p_i\ket{i}\!\bra{i}_{S} \bigotimes_{k=1}^{N_{E}} \tilde{\rho}_l^{(i)},
\end{equation}
where $p_{i} = \bra{i}\rho_{S,0} \ket{i}_{S}$, and where we have used a tilde to denote a state of a sub-environment, rather than an observer system. Note that despite its appearance, $\rho_{\text{eq}}$ in Eq.~\eqref{rho_inf} is not an SBS state, as the distinguishability condition, i.e. $\tilde{\rho}_{l}^{(i)}\tilde{\rho}_{l}^{(j)}=0$ for all $i \neq j$ and all $l$, is not met. Each sub-environment $l$, has a different equilibrium state $\tilde{\rho}_{l}^{(i)}$, conditional on the system's state, indexed by $i$:
\begin{equation} \label{cond_state}
    \tilde{\rho}_{l}^{(i)} = \sum_{n_{l}=1}^{d_l} \bra{E_{n_{l}}^{(i)}}\tilde{\rho}_{l,0}\ket{E_{n_{l}}^{(i)}} \ket{E_{n_{l}}^{(i)}}\!\bra{E_{n_{l}}^{(i)}}.
\end{equation}
The state $\tilde{\rho}_{l}^{(i)}$ is in general mixed, and only depends on the initial state $\tilde{\rho}_{l,0}$ and on the conditional Hamiltonian of the $l^\text{th}$ sub-environment corresponding to the outcome $i$. Note that if the sub-environments were initially correlated, this would not be the case. 

Now, as illustrated in Fig.~\ref{fig:schematic}, we identify sub-environments with observer systems, either one-to-one ($\rho_{k}^{(i)}=\tilde{\rho}_{l}^{(i)}$ for some $k$,$l$), or by grouping sub-environments together into a coarse-grained~\footnote{While the grouping of environments increases the number of possible observables and hence in principle the overall amount of information available (just as zooming out on a map increases the visible landscape), most of the observables on this larger space do not resolve microscopic information about individual components of the observer system. In particular, because the same information (i.e.\ regarding the measurement basis of the system of interest) is now to be encoded into this larger space, the resolving power is diminished and it becomes harder to discern ``where'' this information is stored.} observer system: $\rho_{k}^{(i)}=\bigotimes_{l}\tilde{\rho}_{l}^{(i)}$ where the tensor product is taken over all $l$ associated with that particular $k$. This allows us to compare the effect of increasing dimensionality in both cases. As we shall see later, the choice of identification (i.e. coarse-grained or one-to-one) has a demonstrable effect on the objectivity of the equilibrium state.

The average of the measurement error $\mathcal{E}_{k}$ over the GUE is bounded by the average of the objectivity and equilibration error terms in Eq.~\eqref{error} since all terms are positive, i.e.
\begin{equation} \label{error_bound_GUE}
    \left\langle \mathcal{E}_{k} \right\rangle_{\text{GUE}} \leq \left\langle \mathcal{E}_{\text{obj}} \right \rangle_{\text{GUE}} + \left\langle \mathcal{E}_{\text{eq}} \right \rangle_{\text{GUE}}.
\end{equation}
We numerically approximate this by drawing each conditional Hamiltonian $H_{l}^{(i)}$ from the GUE and averaging over many instances. In the following sections, we will study the two terms $\left\langle \mathcal{E}_{\text{obj}} \right \rangle_{\text{GUE}}$ and $\left\langle \mathcal{E}_{\text{eq}} \right \rangle_{\text{GUE}}$ for different initial states of the environment: pure, maximally mixed and finite-temperature. When both terms approach zero, the average error likewise approaches zero. Our analysis will answer two questions: do systems \textit{in general} equilibrate under chaotic conditional dynamics, and is the equilibrium state objective according to the objectifying observables?

Firstly, we wish to investigate when $\left\langle \mathcal{E}_{\text{eq}} \right \rangle_{\text{GUE}}$ is minimised. As noted above, the conditional Hamiltonians are non-degenerate, and moreover the probability of any energy gap occurring more than once is zero, as is the case for any reasonable chaotic many-body Hamiltonian~\cite{linden2009quantum}. We begin by simplifying $d_{\mathrm{eff}}=1/\operatorname{tr}(\rho_{\text{eq}}^2)$. Using Eq.~\eqref{rho_inf}, we can write the effective dimension in terms of the eigenstates $\{\ket{E_{n_{l}}^{(i)}}\}_{i,l}$ of the conditional Hamiltonians and the initial sub-environment states $\tilde{\rho}_{l,0}$:
\begin{align}
    d_{\mathrm{eff}} = \left[\sum_{i=1}^{d_{S}}p_{i}^{2} \prod_{l=1}^{N_E}\sum_{n_l =1}^{d_l}\bra{E_{n_l}^{(i)}} \tilde{\rho}_{l,0} \ket{E_{n_l}^{(i)}}^{2}  \right]^{-1}. \label{eq:simplified_deff}
\end{align}
We define 
\begin{equation} \label{random_variable}
    X_{n_{l}}^{i} \equiv \bra{E_{n_l}^{(i)}} \tilde{\rho}_{l,0} \ket{E_{n_l}^{(i)}}^{2}
\end{equation}
as a random variable for simplicity of notation. If we assume the system $S$ is initially in an equal superposition (which is, in a sense, the hardest case to distinguish), then we have that $p_{i}=\frac{1}{d_{S}}$ for all $i$. Then, $\left\langle \mathcal{E}_{\text{eq}} \right \rangle_{\text{GUE}}$ reduces to:
\begin{align} \label{eq:expected_val_E_eq}
    \left\langle \mathcal{E}_{\text{eq}} \right \rangle_{\text{GUE}} &= \frac{1}{4}\left\langle\sqrt{ \sum_{i=1}^{d_{S}} \prod_{l=1}^{N_E}\sum_{n_l =1}^{d_{l}} X_{n_{l}}^{i} }\right\rangle_{\text{GUE}}.
\end{align}

\begin{figure*}[t!]
\centering\includegraphics[width=1.0\textwidth]{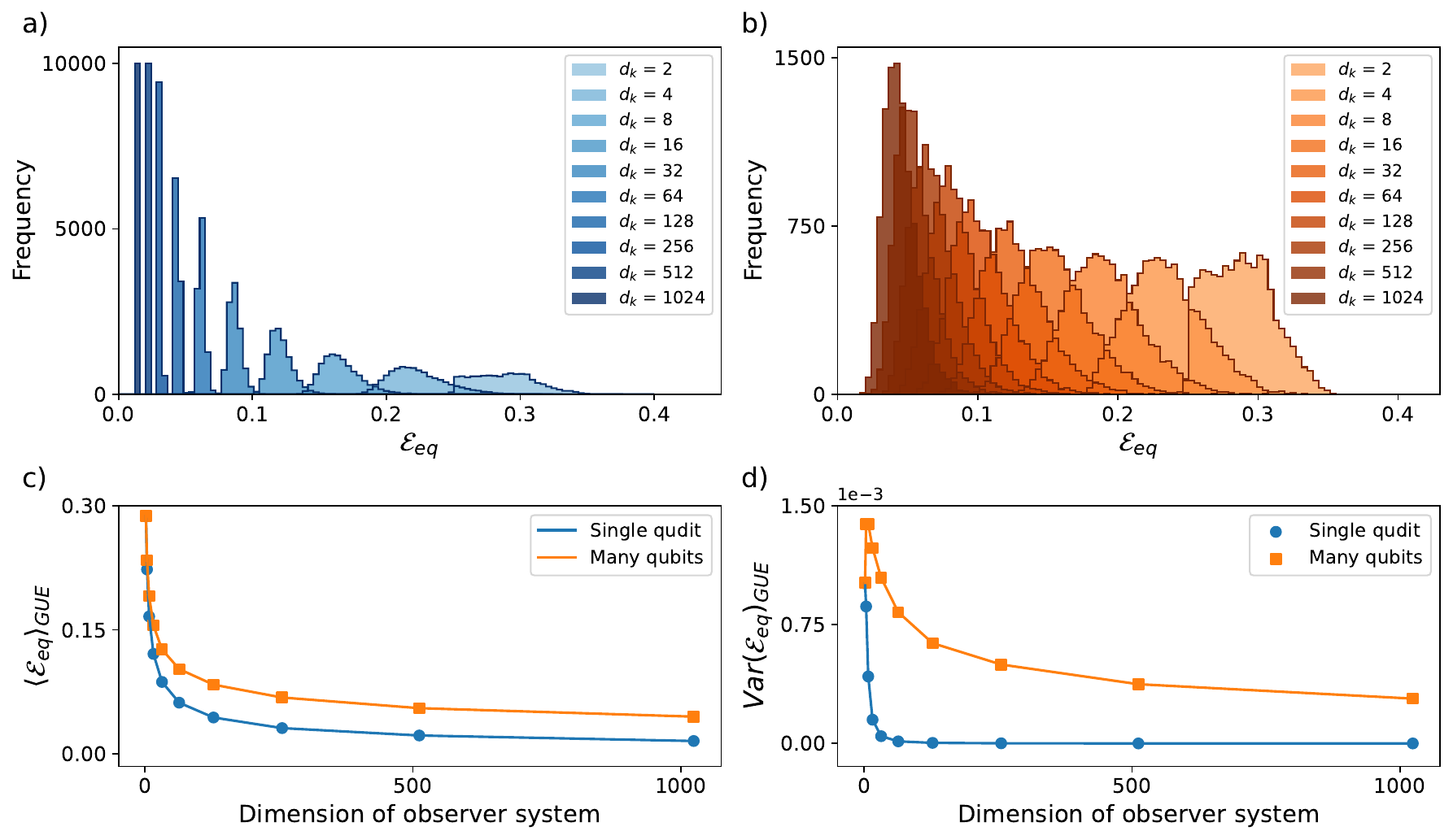}
\caption{Numerical simulation of $ \mathcal{E}_{\text{eq}} $, defined in Eq.~\eqref{eq:D_term}, when sampled over the GUE. Each sample in histograms (a) and (b) considers an initially pure and uncorrelated system-environment state, evolving according to conditional Hamiltonians drawn from the GUE. There are 10,000 samples for each considered observer system dimension $d_{k}$. (a) A single qudit observer system of dimension $d_{k}$ (the scenario in Fig.~\ref{fig:schematic}a). (b) A coarse-grained observer system of $n$ qubits with total dimension $d_{k} = 2^n$ (the scenario in Fig.~\ref{fig:schematic}b). In (c) we plot the average over the GUE of each histogram in single qudit and many-qubit observer system cases as a function of $d_{k}$. In (d) we plot the variance over the GUE. }
\label{fig:effective_dimension}
\end{figure*}

Next, we analyse the average over the GUE of $\mathcal{E}_{\text{obj}}$, where the latter was defined in Eq.~\eqref{eq:F_term}. This term quantifies the (lack of) objectivity of the equilibrium state. When it is zero, $\rho_{\text{eq}}$ is an SBS state and therefore objective. As above, we assume that $p_{i}=\frac{1}{d_{S}}$ for all $i$. Averaged over the GUE, all the fidelity terms in Eq.~\eqref{eq:F_term} are the same, and so may be replaced by any specific term in the summation, e.g.\ $i=0$ and $j=1$, without loss of generality. Noting that the total number of terms in the summation is $d_{S}(d_{S}-1)$, we then have 
\begin{equation}
    \left\langle \mathcal{E}_{\text{obj}} \right \rangle_{\text{GUE}} = (d_{S}-1)  \left\langle F\left(\rho_k^{(i)}, \rho_k^{(j)}\right) \right\rangle_\text{GUE} + \langle \Delta P \rangle_{\text{GUE}}.
\end{equation}
The probability that $F\left(\rho_k^{(i)}, \rho_k^{(j)}\right)=0$ for any random draw from the GUE is zero, as there exists no Hamiltonian for which the equilibrium state has this property exactly~\cite{schwarzhans2025quantum}. As noted above $F\left(\rho_k^{(i)}, \rho_k^{(j)}\right)$ only asymptotically approaches zero when larger and larger observer systems of the environment (so-called macrofractions) are considered.

We study the behaviour of the upper bound in Eq.~\eqref{error_bound_GUE} as the observer systems dimensions $d_{k}$ increase. Specifically, we investigate whether both terms $\left\langle \mathcal{E}_{\text{obj}} \right \rangle_{\text{GUE}}$ and $\left\langle \mathcal{E}_{\text{eq}} \right \rangle_{\text{GUE}}$ approach zero in the limit of very large $d_{k}$. Due to the strong independence-preserving structure of the broadcasting Hamiltonian (i.e. no interactions between sub-environments), we only need to simulate a single sub-environment, and the result can then be used to calculate the error for composite observer systems. We consider two distinct scenarios, illustrated in Fig.~\ref{fig:schematic}. In Fig.~\ref{fig:schematic}a, we consider an observer system comprised of a single qudit with dimension $d_{k}$. In Fig.~\ref{fig:schematic}b, we consider an observer system comprised of $n$-qubit sub-environments, and thus with total dimension $d_{k}=2^n$. Both scenarios are investigated in the simulations in the following sections.

We consider pure, maximally-mixed and thermal initial states. We restrict ourselves to measuring a qubit system $d_{S}=2$, initially in an equal superposition in the measurement basis, such that $\rho_{S,0}=\ket{\psi_{S,0}}\bra{\psi_{S,0}}$, where $\ket{\psi_{S,0}} = \frac{1}{\sqrt{2}}(\ket{0}+\ket{1})$ and $\lbrace \ket{0},\ket{1}\rbrace$ is the measurement basis. In this case, the only distinguishability condition to consider in Eq.~\eqref{eq:F_term} is the function $F\left(\rho_k^{(0)}, \rho_k^{(1)}\right)$ and $\Delta P=0$ as the optimal measurement when distinguishing between two states is a projective measurement~\cite{barnett2009quantum}. For each simulation round, we generate conditional Hamiltonians from the GUE, in which the matrix elements are complex normal random numbers ($\mu=0$ and $\sigma=1$). This is done with the python package TeNPy \cite{tenpy}. We then use the QuTiP package \cite{johansson2012qutip} to numerically calculate $\rho_{\text{eq}}$ and from this $d_{\text{eff}}$ and $F\left(\rho_{k}^{(0)},\rho_{k}^{(1)}\right)$. For higher system dimensions ($d_S >2$), we would expect the behaviour we observe to be broadly similar, since it is primarily driven by the discrepancy between the number of measurement outcomes to be distinguished and the dimensionality of the space into which they are to be encoded, with the latter always being much larger than the former when the environment is reasonably large. The term $\Delta P$ will no longer be identically zero but we expect that as $F(\rho_k^{(i)}, \rho_k^{(j)}) \rightarrow 0$, $\Delta P \rightarrow 0$ as well. In the limit of $F(\rho_k^{(i)}, \rho_k^{(j)})=0$, we know that $\Delta P=0$, yet it remains to be seen exactly how the behaviour of $\Delta P$ changes as the fidelity term approaches zero. This is related to the wider problem of optimising over projective measurements versus POVM's, which has been studied in detail in the context of SBS states in~\cite{acevedo2024spectrum}.

\begin{figure*}[t!]
\centering\includegraphics[width=1.0\textwidth]{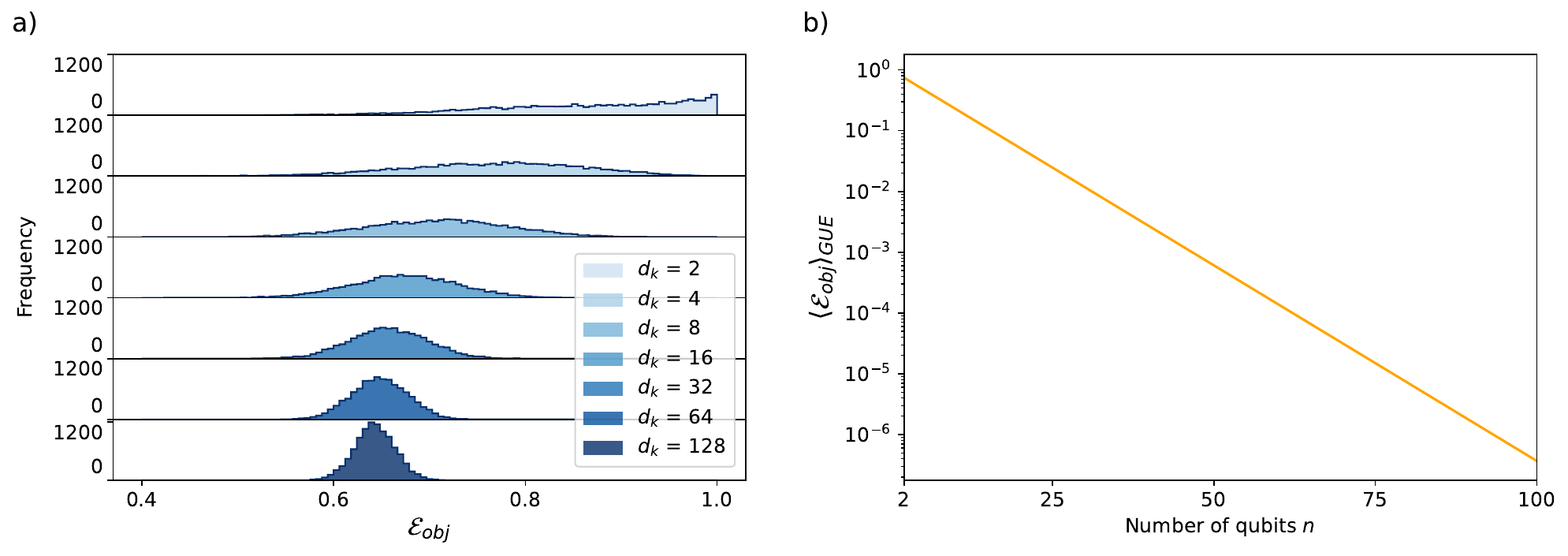}
\caption{ Numerical simulation of $ \mathcal{E}_{\text{obj}} $, defined in Eq.~\eqref{eq:F_term}, via sampling over the GUE. (a) shows histograms for a single qudit observer system (the scenario in Fig.~\ref{fig:schematic}a). Each sample considers an initially pure and uncorrelated system-environment, evolving according to conditional Hamiltonians drawn from the GUE. There are 10,000 samples for each considered dimension $d_{k}$. In (b) we show $\left\langle \mathcal{E}_{\text{obj}} \right \rangle_{\text{GUE}}$ for a coarse-grained observer system consisting of $n$ qubits (the scenario in Fig.~\ref{fig:schematic}b). We present the average fidelity on a log scale. Here we show the necessity of coarse graining in order to approach zero fidelity in the large observer system limit. }
\label{fig:fidelities}
\end{figure*} 

\subsubsection{Pure initial observer states} \label{section_pure_deff}

First, we consider an environment of uncorrelated pure states, such that $\tilde{\rho}_{l,0}=\ket{\psi_{l}}\bra{\psi_{l}}$ for each sub-environment $l$, i.e.~a temperature-zero environment. We analyse both terms in our error bound separately, beginning with $\mathcal{E}_{\text{eq}}$ for pure initial states. The random variable defined in Eq.~\eqref{random_variable} can then be simplified to $X_{n_{l}}^{i}
 =  | \langle E_{n_l}^{(i)} | \psi_{l}\rangle |^{4}$. For our simulations, we can choose any initial pure state, since both $\left\langle \mathcal{E}_{\text{obj}} \right \rangle_{\text{GUE}}$ and $\left\langle \mathcal{E}_{\text{eq}} \right \rangle_{\text{GUE}}$ are averaged over the GUE and so are invariant under unitary transformation -- see Appendices~\ref{sUnitInvariance} and ~\ref{sUnit_invar_eff_dim}. Thus, without loss of generality, we choose $\tilde{\rho}_{l,0} = \ket{0}\bra{0}$.

Fig.~\ref{fig:effective_dimension}a shows histograms (each with 10,000 random samples) of $\mathcal{E}_{\text{eq}}$ for a qubit system, coupled to a single qudit observer system with dimension $d_{k}$ (i.e.~the scenario described in Fig.~\ref{fig:schematic}a).
Fig.~\ref{fig:effective_dimension}b shows the corresponding set of histograms for a qubit system coupled to an observer system of $n$ initially uncorrelated qubits of total dimension $d_{k}$, i.e.~with initial state $\bigotimes_{l=1}^{n} \ket{0}\bra{0}_{l}$ (the scenario described in Fig.~\ref{fig:schematic}b). 

It is expected that our system will typically approach equilibrium (corresponding to a small $\left\langle \mathcal{E}_{\text{eq}} \right \rangle_{\text{GUE}}$) as the total dimension of the observer system grows. Here, we corroborate that hypothesis, indicated by the histograms in Figs.~\ref{fig:effective_dimension}a and \ref{fig:effective_dimension}b tending towards zero as $d_{k}$ increases. There are, however, certain complications: there is a dependence on the Hilbert space structure of the observer system and the corresponding dynamics. Fig.~\ref{fig:effective_dimension}c plots the averages of each histogram from Figs.~\ref{fig:effective_dimension}a and \ref{fig:effective_dimension}b on the same plot. It shows that a single-qudit observer system equilibrates better (on average) in this setting than a many-qubit observer system of the same dimension. We also see from Fig.~\ref{fig:effective_dimension}d that the variance of $\mathcal{E}_{\text{eq}}$ for a single-qudit observer decreases considerably faster with dimension than in the case of the equivalent many-qubit observer system. Each qubit in the many-qubit case interacts individually with the central system in a star-like structure, resulting in very different dynamics to the single-qudit case. This result highlights the dependence of the effective dimension on the form of the interaction. 

\begin{figure*}[t!]
\centering\includegraphics[width=1.0\textwidth]{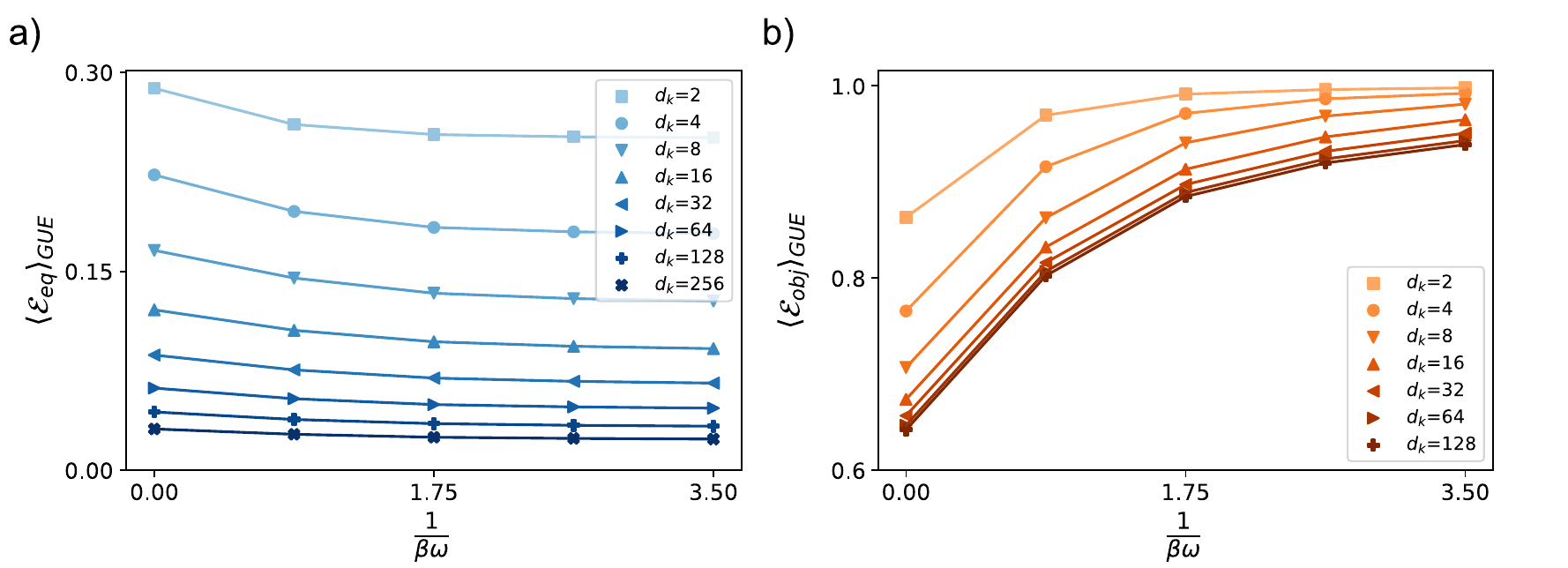}
\caption{ (a) shows $\left\langle \mathcal{E}_{\text{eq}} \right \rangle_{\text{GUE}}$ for a single-qudit observer system with dimension $d_{k}$ (the scenario in Fig.~\ref{fig:schematic}a), that is initially thermal at temperature $T=1/ \beta$ and energy gap $\omega$. (b) shows $\left\langle \mathcal{E}_{\text{obj}} \right \rangle_{\text{GUE}}$ for the same thermal initial states as in (a). Each data point in (a) and (b)  is an average of 10,000 samples over the GUE. }
\label{fig:thermal_states}
\end{figure*}

We now examine the objectivity term in the error bound. Fig.~\ref{fig:fidelities}a shows histograms (10,000 samples each) of $\mathcal{E}_{\text{obj}}$ for single-qudit observer systems of increasing dimension (the scenario in Fig.~\ref{fig:schematic}a). We can see that even for a high-dimensional observer system ($d_{k}=128$), the mean fidelity remains greater than $0.6$. This result may be counter-intuitive, as the overlap between two randomly drawn states decreases with the increase of the Hilbert space dimension (see e.g.\ Section~7.6 of~\cite{Bengtsson2006}). However, this comparison is not quite correct, as we are not simply comparing the overlaps of random states, but rather the result of applying two randomly-drawn pinching maps (defined in Appendix~\ref{sUnitInvariance}, and equivalent to taking the infinite-time average) to $\tilde{\rho}_{l,0} = \ket{0}\bra{0}$. The pinching map in each case depends on eigenvectors of randomly drawn Hamiltonians. We show in Appendix~\ref{sUnitInvariance}, that in the limit of $d_{k} \rightarrow \infty$, the fidelity between conditional states, averaged over the GUE, approaches $0.62$. This has an important implication for the emergence of objectivity. Specifically, it implies that extracting measurement statistics from a single, chaotic high-dimensional qudit observer system generically results in a non-negligible error.

It was shown in~\cite{schwarzhans2025quantum} that coarse graining sub-environments into multipartite observer systems reduces this error. However, our result indicates that in the scenario considered, this coarse graining is not just beneficial, but in fact \textit{necessary} for the emergence of objectivity via equilibration on average. Indeed, in Fig.~\ref{fig:fidelities}b we consider the result of coarse graining $n$ qubits into one observer system (the scenario in Fig.~\ref{fig:schematic}b). We see that $\left\langle \mathcal{E}_{\text{obj}} \right \rangle_{\text{GUE}}$ exponentially approaches zero with respect to the number of qubits (plotted here on a logarithmic scale).

\subsubsection{Maximally-mixed initial states}

At the opposite end of the spectrum, we consider initial sub-environment states that are maximally mixed, i.e.\ at infinite temperature. Let us briefly relax the assumption that the system $S$ is a qubit, and assume instead that it is in an equally-weighted superposition of basis states with an arbitrary dimension. We find that this choice of initial state maximises the effective dimension, i.e.\ $d_{\mathrm{eff}} = d_{S} \cdot (d_{l})^{N_{E}}$ -- see Appendix~\ref{SM:max_mixed_initial_state}. Thus the equilibration error decreases as either $d_{l}$ or $N_E$ increase, and as either tends to infinity then $\mathcal{E}_{\text{eq}}  = \sqrt{d_S}(4\,  d_l^{N_E/2})^{-1}\rightarrow 0$. Indeed, no matter which Hamiltonian is drawn from the GUE, the state of each sub-environment is maximally mixed throughout the evolution and so $\rho_{l}^{(i)}=\tilde{\rho}_{l,0}$ for all $i$, $l$. In this sense, the dynamics decoheres the system being measured, but the initial state $\tilde{\rho}_{l,0}$ is already in local equilibrium and so no information is being exchanged between the system being measured and the environment during the dynamics. (Thus no information is exchanged with the observer systems, too.) Consequently, this initial state leads to a trivial $ \mathcal{E}_{\text{obj}} $, as $F(\rho_{k}^{(i)},\rho_{k}^{(j)})=1$ for any choice of grouping into an observer system $k$. Therefore, for a system initially in an equal superposition in the measurement basis such that $p_{i}=1/d_{S}$ for all $i$, the error bound reduces to:
\begin{align*}
    \mathcal{E}_{k} \leq (d_{S}-1) + \frac{1}{4}\sqrt{\frac{d_{S}}{d_{l}^{N_E}}},
\end{align*}
for any choice of $k$. Noting that the right-hand side is always greater than unity, we see that in this case the bound does not constrain the error in any way and so is not of any use. This is to be expected, since the entire environment, and in particular any observer system, contains no information about the measurement outcome.

\subsubsection{Finite-temperature initial states}

So far, we have considered two opposing examples of initial states: pure and maximally mixed, i.e.~zero and infinite-temperature respectively. For completeness, in this section, we study the intermediate scenario, i.e.\ sub-environments initially at temperatures $0<  T < \infty$. 

A thermal state is defined as
\begin{align}
    \rho^{th} = \frac{e^{-\beta H_{th}}}{Z},
\end{align}
where $Z = \operatorname{tr}(e^{-\beta H_{th}})$ is the partition function, $\beta = 1/k_{B}T$ is the inverse temperature (we set Boltzmann's constant $k_{B}=1$ in all simulations) and $H_{th}$ is some Hamiltonian. Here, we consider an initial state $\rho(0) = \rho_{S,0} \otimes_{l=1}^{N_E} \tilde{\rho}_{l}^{th}$, where we again take the system $S$ to be a qubit, initially in an equal superposition in the measurement basis, and where $\tilde{\rho}_{l}^{th}$ is a thermal state of the $l^{\text{th}}$ sub-environment. Here we only consider the case where each of the observer systems $k$ consist of a single sub-environment, i.e.\ a single qudit of dimension $d_{k}=d_{l}$ (this is the scenario in Fig.~\ref{fig:schematic}a). We show in Appendix~\ref{SM:thermal} that the error bound, when averaged over the GUE, is invariant under unitary transformations of $H_{th}$. Therefore, our choice of eigenstates in $H_{th}$ will not affect our simulation results. We assume $H_{th}$ has equally-spaced eigenvalues and note that in this case, a different choice of eigenvalues is equivalent to an overall rescaling, and therefore to a different temperature. We make use of a built-in QuTiP function (\texttt{qutip.enr\_thermal\_dm}) to generate our thermal initial states in the excitation-restricted number space at inverse temperature $\beta = 1/k_{B}T$ \cite{johansson2012qutip}. For details on our choice of $H_{th}$, see Appendix~\ref{SM:thermal_num}.

In Fig.~\ref{fig:thermal_states}a, we show the obtained values of the equilibration term $\left\langle \mathcal{E}_{\text{eq}} \right \rangle_{\text{GUE}}$ for increasing initial sub-environment temperature. We see that in general, as the temperature of the initial environment state increases, $\left\langle \mathcal{E}_{\text{eq}} \right \rangle_{\text{GUE}}$ decreases. However, the effect of temperature becomes less impactful as the observer system dimension increases. These observations accord with the intuition that the a higher temperature will lead to a failure to encode information well, rather than failure to equilibrate.

In Fig.~\ref{fig:thermal_states}b, we analyse $\left\langle \mathcal{E}_{\text{obj}} \right \rangle_{\text{GUE}}$ for the same scenario and see that $\left\langle \mathcal{E}_{\text{obj}} \right \rangle_{\text{GUE}}$ increases, tending to one, as we increase the temperature of the initial environment. This indicates that higher temperature environments are less able to exchange information with the system $S$, as expected. Loosely speaking, mixed-state environments are noisy and so less able to acquire and store information about the system of interest. This was previously investigated in the context of Quantum Darwinism \cite{zwolak2009quantum,12RiedelZurekZwolak} and further seen for spectrum broadcast structure states in \cite{mironowicz2017monitoring}. Crucially, however, Fig.~\ref{fig:thermal_states}b highlights the interplay between temperature and dimensionality, showing that increasing dimensionality protects against the noise-inducing effect of temperature over all temperature ranges.

Our results suggest that the broadcasting of information via closed-system equilibration may overcome the noisiness of finite-temperature environments as the dimensionality of the system becomes macroscopic, an essential prerequisite for the former to be a realistic model of the measurement process. We note however that this analysis is limited by the small size of our largest system relative to a realistic many-body quantum system, which precludes us from making a definitive statement about the macroscopic limit.

\section{Discussion \& Conclusions}

When an observer measures a quantum system of interest, information about a particular observable is extracted. This is modelled mathematically as the observer applying an observable operator to the system being measured and obtaining a measurement outcome that corresponds to an eigenstate of the observable. In a dynamical measurement model, when the system is measured, information in its pointer basis is exchanged with the environment via their interaction. 

Building on previous work~\cite{schwarzhans2025quantum}, we assert the hypothesis that this process is unitary, and correspond to a spontaneous, uncontrolled, irreversible process of equilibration. This proposal, dubbed the Measurement-Equilibration Hypothesis (MEH), stands in contrast to the conventional von Neumann model of a measurement-type interaction between a system and a detector. The MEH thus frames quantum measurements as an inherently thermodynamic (or more precisely, statistical mechanical) phenomenon that directly results from the universe's tendency towards entropy maximisation. Not only does the large environment explain the emergence of \textit{irreversibility}, it also explains the emergence of \textit{objectivity}, since one or multiple observer systems can access information about the measured system in this scenario, and they will agree on the outcome of a particular measurement. We note that in Ref.~\cite{busch1996quantum} a different definition of objectivity is discussed in the context of measurements. Here, we focus on the definition of objectivity as presented in the quantum Darwinism literature~\cite{zurek2009quantum}. 

In this work, we provided tangible process towards quantifying specific parts of the MEH. We do so by defining the set of \textit{objectifying observables}, which optimally encode measurement statistics from the system being measured. For example, in a Stern-Gerlach experiment, the objectifying observables would be the position of the particles, with the spontaneous, uncontrolled equilibration happening when the particles interact with the photographic plate at the screen. From this position observable, we learn the measurement outcome (the spin) of the particle. Applying this to the equilibrium state, we solve the question of how macroscopic, equilibrating observables can encode the measurement outcome -- a question whose answer is vital for the tenability of the MEH.

We then constructed an error bound on the measurement with respect to the objectifying observables. We use numerical methods to find that this error decreases as the environment dimension increases and importantly, when the environment is coarse-grained. These results indicate the objectifying observables readily equilibrate for large Hilbert spaces of the environment. 

Coarse graining has been shown to play a role in the emergence of objectivity in a noisy photonic environment~\cite{korbicz2014objectivity}, for example, as well as in thermodynamic investigations of the measurement process~\cite{schwarzhans2025quantum}. Interestingly, our numerical studies highlight that the emergence of objectivity necessarily requires coarse graining of observers into observer systems in some cases. Without coarse graining, the error bound remains high, even when the dimension of the environment is large. This is well-motivated by a realistic view of how a detector interacts with a quantum system -- it reflects the fact that not every quantum degree of freedom of the detector needs to encode information about the system for the detector as a whole to be able to record the measurement outcome.

We have shown that under dynamics described by the conditional Hamiltonian structure (i.e.\ broadcasting Hamiltonians), the objectifying observables generically encode the measurement statistics and equilibrate. However, we have not shown that \textit{all} observables that equilibrate encode the measurement statistics. This is intuitive, as we would not expect every degree of freedom to perfectly record measurement outcomes. For example, in a Stern-Gerlach experiment, both the particle position and momentum equilibrate at the end of the experiment, but only one of those two observables, the position, encodes the spin state of the system \cite{busch1996quantum}. 

Here, we also emphasise that the method used to construct the objectifying observables is independent of dynamics, meaning our method applies beyond the conditional Hamiltonians considered in this work. However, because of this independence, one can construct example dynamics that result in observables that are maximally incapable of reproducing the measurement statistics of the system. Hence, it is important to note that the observables are not unconditionally objectifying and their ability to objectify depends heavily on the dynamics.

Of course, the class of measurements considered here was somewhat restrictive. Extensions of this work to more general classes of POVMs and to sequential (potentially non-commuting) measurements are ongoing. Generally speaking, the properties of the MEH equilibration process will depend on the structure of the interaction Hamiltonian, particularly its system factor, but we do not anticipate the main results presented here to be drastically different when applied to the case of POVMs.

Our work rests on the application of closed-system equilibration to a framework which assumes that the situation post-measurement can be associated with a quantum state, arrived at through some overall unitary evolution, and without necessarily selecting a given outcome. Closed-system equilibration is a well-established phenomenon (see~\cite{gogolin2016equilibration} and references therein), but the assumption of unitary evolution without selecting a given outcome, while widespread and consistent with experiment~\cite{dyte2024wave}, is not universally adopted. It may be contrasted, for example, with the hypothesis of objective wavefunction collapse, where it is proposed that the dynamics have an inherently stochastic component~\cite{bassi2023collapse}.

Many aspects of quantum measurements are still to be investigated in the MEH paradigm. The paradigm provides tools to explore measurement timescales and the relationship between measurement speed and observer size. Investigating the process of entropy maximisation, particularly the entropies associated with observables~\cite{meier2025emergence}, is also of interest. The examination of more complicated many-body systems, such as many-body fermion or boson chains, would bring the MEH closer to real-world phenomena and potential experimental proposals. Lastly, a generalisation effort could extend the paradigm to continuous variables, expanding the range of physical degrees of freedom the MEH model can be applied to.

\section{Acknowledgements}

The authors thank Emanuel Schwarzhans, Will McCutcheon, Felix Binder, Jaros\l{}aw Korbicz, Kl\'{a}ra Baksov\'{a} and Marcus Huber for helpful discussions, as well as Alberto Acevedo for noticing and assisting in correcting a small mistake in the appendices. M.~P.~E.~L. thanks Yuri Minoguchi for many useful and interesting conversations about random matrix theory. This work was supported by the EPSRC grant EP/SO23607/1, the European Research Council (ERC) Starting grant PIQUaNT (950402), Austrian Science Fund (FWF) grant TOMAIQS (ESP7464924), and the ERC Consolidator grant Cocoquest (101043705). SW acknowledges funding from the European Union’s Horizon 2020 research and innovation programme under the Marie Skłodowska-Curie grant agreement No 89224. This publication was made possible through the support of Grant 62423 from the John Templeton Foundation. The opinions expressed in this publication are those of the authors and do not necessarily reflect the views of the John Templeton Foundation. The authors acknowledge TU Wien Bibliothek for financial support through its Open Access Funding Programme.

\section{Data Availability}

The data that support the findings of this article are openly available~\cite{DataAvailability}. 

\newpage
\bibliographystyle{unsrt}

\newpage

\appendix

\widetext

\section{Finding optimal projectors and a close candidate SBS state} \label{SM:candidate_SBS}
Using a convex optimisation method, for each observer system, we can find a set of optimal projectors to distinguish between the states of the measured system $S$. Given a system-environment Hilbert space $\mathcal{H}_{S}\otimes \mathcal{H}_{E}$, with $N_E$ sub-environments, such that $\mathcal{H}_{E}=\mathcal{H}_{1}\otimes \dots \otimes \mathcal{H}_{N_{E}}$, one can fix the pointer basis of $S$. Then, with a general equilibrium state written in this pointer basis, we can split the density matrix into diagonal and off-diagonal blocks:
\begin{align} \label{eq:general_state_seperated}
    \rho_{eq}=  \sum_{i}^{d_{S}} q_{i}|i\rangle\langle i|_{S} \otimes \rho^{(i)} + \gamma_{SE},
\end{align}
where $\gamma_{SE}$ is the off-diagonal part. This density matrix is completely independent of dynamics. The only assumption made is that of the pointer basis of the system $\{\ket{i}_S\}_{i}$, which we assume is optimal (for the case of our considered Hamiltonian, Eq.~\eqref{ham}, this is true when averaging over the GUE). We then assume there exist projective measurements to distinguish between the `branches' of the environment $\rho^{(i)}$ and $\rho^{(j)}$ for all $i \neq j$ (this is not the case if $\rho^{(i)}=\rho^{(j)}$). 

For simplicity, we will start by considering a single observer performing a projective measurement over the entire environment, generalising to multiple observer systems later. The branches of an SBS state have non-overlapping support and so can be distinguished perfectly with a single projective measurement. To find how close our state is to an SBS state, we find the set of projectors $\{\Pi^{i}\}_{i}$ that gives the largest probability of success when measuring the state $\rho^{(i)}$ for all $i$ (ignoring the off-diagonal part $\gamma_{SE}$ for now). We can view this as a state discrimination problem, in which we have a set of states $\{ 
\rho^{(i)} \}_{i}$, each state occurring with probability $q_{i}$. We want to distinguish between the states with the minimum average error~\cite{montanaro2008lower,barnum2002reversing}. We use the Python library CVXPY \cite{diamond2016cvxpy,agrawal2018rewriting} to perform the maximisation problem defined in the main text as the probability of success when optimising over projectors (Eq.~\eqref{prob_succ}) \cite{le2018objectivity,mironowicz2017monitoring}:
\begin{align}
     P_{\{\Pi^i\}}^{\text{Succ}} = \max_{\{\Pi^{i}\}_i} \sum_{i} q_{i} \operatorname{tr}\left(\rho^{(i)}\Pi^{i}\right).
\end{align}
Using the set of optimal projectors found via the optimisation, we define a candidate SBS state:
\begin{align} \label{eq:candidate_SBS_one_observer}
    \rho^{\text{SBS}}= \sum_i q_i    |i\rangle\langle i|_{S} \otimes \sigma^{(i)},   \quad \sigma^{(i)} = \frac{\Pi^i }{ \operatorname{tr}\left[\Pi^i\right]},
\end{align}
where the probabilities $q_{i}$ corresponding to each pointer state $\ket{i}_S$ are the same as the initial probabilities in Eq.~\eqref{eq:general_state_seperated}. This state varies slightly from the state constructed in \cite{mironowicz2017monitoring}, where new probabilities are defined. It is important that any measurement process preserves the outcome probabilities and therefore any SBS state we are constructing must also preserve the probabilities $\{q_{i}\}_{i}$.

Now we will generalise to multiple observer systems. Let's say we have $K$ observer systems within the environment, such that $K \leq N_{E}$ and we index each observer system $k$. Again, we begin with a general state, which by fixing the pointer basis of $S$, we can write in the form of Eq.~\eqref{eq:general_state_seperated}. To construct the probability of success, we again ignore the off-diagonal section $\gamma_{SE}$ and define 
\begin{align}
    \rho_{k}^{(i)} = \operatorname{tr}_{k' \neq k} \left[ \rho^{(i)}  \right] = \frac{1}{q_{i}} \operatorname{tr}_{k' \neq k} \left[ \bra{i}\rho_{eq}\ket{i}_{S}  \right].
\end{align}
The search for optimal projectors is performed separately for each observer system $k$, resulting in a set of success probabilities
\begin{equation} \label{prob_succ_SM}
    \left\{P_{\{\Pi_{k}^i\}}^{\text{Succ}} = \max_{\{\Pi_{k}^{i}\}_{i}} \sum_{i} q_{i} \operatorname{tr}\left(\rho_{k}^{(i)}\Pi_{k}^{i}\right) \right\}_{k}. 
\end{equation}
The candidate SBS state becomes:
\begin{align} \label{eEquilProximateSBS}
        \rho^{\text{SBS}} &= \sum_{i} q_{i} \ket{i}\bra{i}_{S} \bigotimes_{k=1}^{K} \sigma_{k}^{(i)},   \quad \sigma_{k}^{(i)} = \frac{\Pi_k^{i}}{ \operatorname{tr}\left[\Pi_k^{i}\right]}
\end{align}
where $\lbrace\Pi_k^{i}\rbrace_{i,k}$ are the optimal projectors found in the maximisation in Eq.~\eqref{prob_succ_SM}.

\section{Derivation of the error bound for objectifying observables} \label{SM_error_derivation}

This section derives the error bound, defined in Eq.~\eqref{error}. This error quantifies the probability that an observer system will not accurately reproduce the measurement outcome. If this error is vanishing, one can conclude that the corresponding dynamics describe a measurement. We emphasise that the derivation of the error bound is independent of considered dynamics. The error is based on the distinguishability of states \cite{short2011equilibration}. This definition quantifies how well a POVM can distinguish between two states. For two states $\rho_{1}$ and $\rho_{2}$, given a measurement (POVM) $M$, the distinguishability of states is:
\begin{equation}
    D_M\left(\rho_1, \rho_2\right) \equiv \frac{1}{2} \sum_i\left|\operatorname{tr}\left(M_i \rho_1\right)-\operatorname{tr}\left(M_i \rho_2\right)\right|,
\end{equation}
where $M$ is described by a positive operator $M_{i}$ for each outcome $i$, such that $\sum_{i} M_{i} = \mathbb{1}$. For each observer, we define an objectifying observable (Eq.~\eqref{obs}) that is constructed from a set of projective measurements found from the method in Appendix~\ref{SM:candidate_SBS}. We likewise use them to construct the equilibrium-proximate SBS state $\rho_{\text{eq}}^{\text{SBS}}$ from a general equilibrium state $\rho_{\text{eq}}$ of the form of Eq.~\eqref{eq:general_state_seperated}, specifically via Eq.~\eqref{eEquilProximateSBS}. For each observer system, indexed $k$, we then look at the distinguishability between $\rho(t)$ and $\rho_{\text{eq}}^{\text{SBS}}$ given the observer applies the objectifying observable $\hat{O}_{k} = \sum_{i=1}^{d_S} c_{i} O_{k}^{i} = \sum_{i=1}^{d_S} c_i \mathbb{1}_{S} \otimes \Pi_{k}^{i} \otimes \mathbb{1}_{k' \neq k}$. Distinguishability obeys the triangle inequality so,
\begin{equation} \label{eq:D_triangle}
    D_{\hat{O}_{k}}\left(\rho(t), \rho_{\text{eq}}^{\text{SBS}} \right) \leq D_{\hat{O}_{k}}\left(\rho(t), \rho_{\text{eq}}\right) + D_{\hat{O}_{k}}\left(\rho_{\text{eq}},  \rho_{\text{eq}}^{\text{SBS}} \right).
\end{equation}
Firstly, looking at the second term on the RHS,
\begin{align}
    D_{\hat{O}_{k}}\left(\rho_{\text{eq}},  \rho_{\text{eq}}^{\text{SBS}} \right) &= \frac{1}{2} \sum_i\left|\operatorname{tr}\left(O_k^{i}  \rho_{\text{eq}} \right) -\operatorname{tr}\left(O_k^{i} \rho_{\text{eq}}^{\text{SBS}} \right) \right|,
\end{align}
we show that
\begin{align}
    \operatorname{tr}\left(O_k^{i}  \rho_{\text{eq}} \right) &= \operatorname{tr}\left[ O_k^{i}  \left(\sum_j q_j |j\rangle\langle j| \otimes \rho^{(j)}+\gamma_{SE}  \right)\right] \notag\\
    & = \operatorname{tr}\left[ O_k^{i}  \left(\sum_j q_j |j\rangle\langle j| \otimes \rho^{(j)}\right)\right] + \operatorname{tr}\left[ O_k^{i}  \gamma_{SE} \right] \notag\\
    & = \operatorname{tr}_S \operatorname{tr}_E  \left[ O_k^{i}  \left(\sum_j q_j |j\rangle\langle j| \otimes \rho^{(j)}\right)\right] +  \operatorname{tr}_E\left[ O_k^{i}  \operatorname{tr}_S \left(\gamma_{SE}\right) \right] \notag\\
    & =  \sum_j q_j \operatorname{tr}_E  \left[ \left( \Pi_{k}^{i} \otimes \mathbb{1}_{k' \neq k}\right)  \left( \rho^{(j)}\right)\right] + 0 \notag\\
    & =  \sum_j q_j \operatorname{tr}_k  \left[ \left( \Pi_{k}^{i} \otimes \mathbb{1}_{k' \neq k}\right) \operatorname{tr}_{k'\neq k} \left( \rho^{(j)}\right)\right]  \notag\\
    & = \sum_{j} q_{j}\operatorname{tr}\left( \Pi_{k}^{i} \rho_{k}^{(j)} \right) \notag\\
    & = q_{i}\operatorname{tr}\left( \Pi_{k}^{i} \rho_{k}^{(i)} \right) + \sum_{j \neq i} q_{j}\operatorname{tr}\left( \Pi_{k}^{i} \rho_{k}^{(j)} \right) 
\end{align}
where we used that $\operatorname{tr}_{k'\neq k}\left(\rho^{(j)}\right)=\rho_{k}^{(j)}$ and that the off-diagonal term has zero trace: $\operatorname{tr}_S \left( \gamma_{SE}\right)=0$. Next, we find
\begin{align} \operatorname{tr}\left(O_k^{i} \rho_{\text{eq}}^{\text{SBS}} \right) &= \operatorname{tr}\left[ \left(\mathbb{1}_{S} \otimes \Pi_{k}^{i} \otimes \mathbb{1}_{k' \neq k}\right)  \left( \sum_{j} q_j  |j\rangle\langle j| \bigotimes_{m=1}^{N_{O}} \sigma_{m}^{(j)}\right)\right] \notag\\
    & = \operatorname{tr}_{E}\operatorname{tr}_{S} \left[ \left(\mathbb{1}_{S} \otimes \Pi_{k}^{i} \otimes \mathbb{1}_{k' \neq k}\right)  \left( \sum_{j} q_j  |j\rangle\langle j| \bigotimes_{m=1}^{N_{O}} \sigma_{m}^{(j)}\right)\right] \notag\\
    & = \sum_{j} q_{j} \operatorname{tr}_{E}\left[ \left( \Pi_{k}^{i} \otimes \mathbb{1}_{k' \neq k}\right)  \left(  \bigotimes_{m=1}^{N_{O}} \sigma_{m}^{(j)}\right)\right] \notag\\
    & = \sum_{j} q_{j} \operatorname{tr}_{k}\operatorname{tr}_{k' \neq k}\left[ \left( \Pi_{k}^{i} \otimes \mathbb{1}_{k' \neq k}\right)  \left(  \bigotimes_{m=1}^{N_{O}} \sigma_{m}^{(j)}\right)\right] \notag\\
    & = \sum_{j} q_{j} \operatorname{tr}_{k}\left[  \Pi_{k}^{i} \sigma_{k}^{(j)}\right] \notag\\
    & = \sum_{j} q_j   \frac{\operatorname{tr}\left(\Pi_{k}^{i} \Pi_k^{j} \right)}{ \operatorname{tr}\left[\Pi_k^{j}\right]}  \notag\\
    & = \sum_{j} q_j   \delta_{ij} \notag\\
    & = q_i,
\end{align}
using that $\sigma_{k}^{(j)} = \Pi_{k}^{j} / \operatorname{tr}\left[\Pi_{k}^{j}\right]$ (defined in Eq.~\eqref{eEquilProximateSBS}). This results in 
\begin{align}
    D_{\hat{O}_{k}}\left(\rho_{\text{eq}},  \rho_{\text{eq}}^{\text{SBS}} \right) &= \frac{1}{2} \sum_{i} \left| q_{i}\operatorname{tr}\left( \Pi_{k}^{i} \rho_{k}^{(i)} \right) + \sum_{j \neq i} q_{j}\operatorname{tr}\left( \Pi_{k}^{i} \rho_{k}^{(j)} \right) - q_i \right| \\
    &= \frac{1}{2} \sum_{i} \left|  \left[ q_i\operatorname{tr}\left( \Pi_{k}^{i} \rho_{k}^{(i)} \right) - q_i\right] + \sum_{j \neq i} q_{j}\operatorname{tr}\left( \Pi_{k}^{i} \rho_{k}^{(j)} \right)  \right| \\
    & \leq \frac{1}{2} \sum_{i} \left[ \left| q_i \operatorname{tr}\left( \Pi_{k}^{i} \rho_{k}^{(i)} \right) - q_i  \right| + \left| \sum_{j \neq i} q_{j}\operatorname{tr}\left( \Pi_{k}^{i} \rho_{k}^{(j)} \right)  \right| \right]\\
    & \leq \frac{1}{2} \left[\sum_i q_i- \sum_{i} q_{i} \operatorname{tr} \left( \rho_{k}^{(i)} \Pi_{k}^{i} \right) \right]+ \frac{1}{2}\sum_{j \neq i} q_{j}\operatorname{tr}\left( \Pi_{k}^{i} \rho_{k}^{(j)} \right) \\
    & = \frac{1}{2}  \left[1 - \sum_{i} q_{i} \operatorname{tr} \left( \rho_{k}^{(i)} \Pi_{k}^{i} \right)\right] + \frac{1}{2}\sum_{j \neq i} q_{j}\operatorname{tr}\left( \Pi_{k}^{i} \rho_{k}^{(j)} \right) , \label{D_bound}
\end{align}
using the property of the modulus: $|x+y| \leq |x| + |y|$ and that $\sum_{i}q_{i}=1$ . Looking separately at the second term in Eq.~\eqref{D_bound} and assuming $d_S=2$ for now, 
\begin{align}
    \frac{1}{2} \sum_{j \neq i} q_{j}\operatorname{tr}\left( \Pi_{k}^{i} \rho_{k}^{(j)} \right) & = \frac{1}{2}\left[ 
 q_0 \operatorname{tr}\left( \Pi_{k}^{1} \rho_{k}^{(0)} \right) + q_1 \operatorname{tr}\left( \Pi_{k}^{0} \rho_{k}^{(1)} \right) \right] \\
 & = \frac{1}{2}\left[ q_0 \operatorname{tr}\left( \left( \mathbb{1}- \Pi_{k}^{0} \right) \rho_{k}^{(0)} \right) + q_1 \operatorname{tr}\left( \left( \mathbb{1}- \Pi_{k}^{1} \right) \rho_{k}^{(1)} \right) \right] \\
 & = \frac{1}{2}\left[ q_0 \operatorname{tr}\left(\rho_{k}^{(0)}\right) -  q_0\operatorname{tr}\left( \Pi_{k}^{0} \rho_{k}^{(0)} \right) +q_1 \operatorname{tr}\left(\rho_{k}^{(1)}\right) -  q_1\operatorname{tr}\left( \Pi_{k}^{1} \rho_{k}^{(1)} \right) \right] \\
 & = \frac{1}{2}\left[  q_0 + q_1 - q_0\operatorname{tr}\left( \Pi_{k}^{0} \rho_{k}^{(0)} \right) - q_1 \operatorname{tr}\left( \Pi_{k}^{1} \rho_{k}^{(1)} \right) \right] \\
 & = \frac{1}{2}\left[  1 - \sum_{i} q_i \operatorname{tr}\left( \Pi_{k}^{i} \rho_{k}^{(i)} \right) \right] \label{B11} ,
\end{align}
which allows us to simplify $D_{\hat{O}_{k}}\left(\rho_{\text{eq}},  \rho_{\text{eq}}^{\text{SBS}} \right)$ to
\begin{align}
    D_{\hat{O}_{k}}\left(\rho_{\text{eq}},  \rho_{\text{eq}}^{\text{SBS}} \right) =  \left[1 - \sum_{i} q_{i} \operatorname{tr} \left( \rho_{k}^{(i)} \Pi_{k}^{i} \right)\right].
\end{align}
The term $\sum_{i} q_{i} \operatorname{tr} \left( \rho_{k}^{(i)} \Pi_{k}^{i} \right)$ is the probability of success in the state discrimination problem \cite{barnett2009quantum}.

For the following, we introduce two types of probabilities of success: the probability of success when using a projective measurement with projectors $\{\Pi_i\}$, which we label $P_{\{\Pi_i\}}^{\text{Succ}}$, and the probability of success when using a POVM with elements $\{E_i\}$, which we label $P_{\{E_i\}}^{\text{Succ}}$. Specifically,
\begin{align}
    P_{\{\Pi_i\}}^{\text{Succ}} &= \sum_{i} q_{i} \operatorname{tr} \left( \rho_{k}^{(i)} \Pi_{k}^{i} \right), \\
    P_{\{E_i\}}^{\text{Succ}} & = \sum_{i} q_{i} \operatorname{tr} \left( \rho_{k}^{(i)} E_{k}^{i} \right).
\end{align}

It was shown in~\cite{montanaro2008lower,barnum2002reversing,mironowicz2017monitoring} that the probability of error, when optimising over all POVMs, is bounded from above (see also \cite{acevedo2024spectrum}):
\begin{align}
    1-P_{\{E_i\}}^{\text{Succ}} = 1 - \sum_{i} q_{i} \operatorname{tr} \left( \rho_{k}^{(i)} E_{k}^{i} \right) \leq \sum_{i \neq j} \sqrt{q_{i} q_{j}} F\left(\rho_{k}^{(i)}, \rho_{k}^{(j)}\right),
\end{align}
where $F(\rho, \sigma)$ is the fidelity. In our case, however, we are restricted to an optimisation over projective measurements, not POVMs. This is due to the distinguishability condition for SBS states. The conditional sub-environment states in $\rho_{\text{eq}}^{\text{SBS}}$ are orthogonal and found by maximising over the set of projectors, as outlined in Appendix~\ref{SM:candidate_SBS}. Yet we can still use this bound by noting that the probability of success when optimising over projectors will always be less than or equal to the probability of success when optimising over all POVMs, i.e. $P_{\{\Pi_i\}}^{\text{Succ}} \leq P_{\{E_i\}}^{\text{Succ}}$. This is because projective measurements are a subset of POVMs. Using this fact, we can write:
\begin{align}
    1 - P_{\{E_i\}}^{\text{Succ}} - P_{\{\Pi_i\}}^{\text{Succ}} \leq \sum_{i \neq j} \sqrt{q_{i} q_{j}} F\left(\rho_{k}^{(i)}, \rho_{k}^{(j)}\right) - P_{\{\Pi_i\}}^{\text{Succ}},
\end{align}
which can be rearranged to obtain:
\begin{align}
    1- P_{\{\Pi_i\}}^{\text{Succ}} \leq \sum_{i \neq j} \sqrt{q_{i} q_{j}} F\left(\rho_{k}^{(i)}, \rho_{k}^{(j)}\right) + \Delta P ,
\end{align}
where we define $\Delta P := P_{\{E_i\}}^{\text{Succ}} - P_{\{\Pi_i\}}^{\text{Succ}} \geq 0$.

Note that in the case of $d_S=2$, there are only two states to be distinguished between, i.e.\ $\rho_{k}^{(0)}$ and $\rho_k^{(1)}$ for each observer system $k$. In this case, the optimal POVM is always a projector~\cite{barnett2009quantum}. This implies that for $d_S=2$, we simply have:
\begin{align}
    D_{O_{k}}\left(\rho_{\text{eq}},  \rho_{\text{eq}}^{\text{SBS}} \right) \leq  \sum_{i \neq j} \sqrt{q_{i} q_{j}} F\left(\rho_{k}^{(j)}, \rho_{k}^{(i)}\right).
\end{align}

Now, we show that \eqref{B11} holds more generally for $d_S>2$. Once again, we begin with the second term in Eq.~\eqref{D_bound}:
\begin{align}
    \sum_{j \neq i} q_{j}\operatorname{tr}\left( \Pi_{k}^{i} \rho_{k}^{(j)} \right) & =  \sum_{j } q_{j}\operatorname{tr}\left( \left( \sum_{i\neq j} \Pi_{k}^{i} \right) \rho_{k}^{(j)} \right) \\
    & =  \sum_{j } q_{j}\operatorname{tr}\left( \left( \mathbbm{1} - \Pi_{k}^{j} \right) \rho_{k}^{(j)} \right) \\
     & = \sum_{j}  q_j  - \sum_{j} \left[q_j \operatorname{tr}\left(\Pi_{k}^{j}  \rho_{k}^{(j)} \right) \right] \\
     & = 1 - \sum_{j}  \left[q_j \operatorname{tr}\left(\Pi_{k}^{j}  \rho_{k}^{(j)} \right) \right], 
\end{align}
where we used the fact that  $\left(\sum_{i\neq j}\Pi_{k}^{i}\right) + \Pi_{k}^{j}= \mathbbm{1}$. For the case where $d_S>2$, it is not necessarily the case that the optimal POVM will be a projector, hence $\Delta P$ may be non-zero and we must include it in our error bound.

For general $d_S$, we therefore have that
\begin{align}
    D_{O_{k}}\left(\rho_{\text{eq}},  \rho_{\text{eq}}^{\text{SBS}} \right) \leq \sum_{i \neq j} \sqrt{q_{i} q_{j}} F\left(\rho_{k}^{(j)}, \rho_{k}^{(i)}\right) + \Delta P
\end{align}

Now we return to the first term on the RHS of the inequality in Eq.~\eqref{eq:D_triangle}. We know that the time average of the first term in the inequality can be bounded~\cite{short2011equilibration}:
\begin{align}
    \left\langle D_{\mathcal{M}}\left(\rho(t), \rho_{\text{eq}}\right) \right\rangle_{\infty} \leq \frac{N(\mathcal{M})}{4 \sqrt{d_{\mathrm{eff}}}},
\end{align}
where $\mathcal{M}$ is a finite set of POVMs such that $D_{\mathcal{M}}(\rho, \sigma)=\max _{M \in \mathcal{M}} D_M(\rho, \sigma)$ and $N(\mathcal{M})$ is the total number of outcomes for all measurement in $\mathcal{M}$. In our cases, the considered measurement set is simply $\mathcal{M} = O_{k}$. Each observer system, indexed $k$, has a set of $d_{S}$ projectors and therefore $d_{S}$ outcomes, such that $N(\mathcal{M}) = d_{S}$.

Overall we can define a time averaged measurement error $\mathcal{E}_{k}$ (as stated in the main text without the $\Delta P$ term):
\begin{equation}
    \mathcal{E}_{k} = \left\langle D_{\hat{O}_{k}}\left(\rho(t), \rho_{\text{eq}}^{\text{SBS}} \right)  \right\rangle_{\infty}\leq   \sum_{i \neq j} \sqrt{q_{i} q_{j}} F\left(\rho_{k}^{(i)}, \rho_{k}^{(j)}\right) + \Delta P +  \frac{d_{S}}{4 \sqrt{d_{\mathrm{eff}}}}. \label{error_SM}
\end{equation} 
This is the error, averaged over all times, for a single observer to determine the state of the measured system, using a single projective measurement on their observer system. For simplicity in the main text we omit the $\Delta P$ term, which we anticipate to be small even in cases where it cannot be assumed to be zero.

\section{Properties of the average fidelity with respect to the Gaussian unitary ensemble}

\subsection{Unitary invariance with respect to the initial state} \label{sUnitInvariance}

Here we show that the average fidelity in Eq.~\eqref{eq:F_term} is invariant under a unitary transformation of the initial state on the sub-environment $l$.

Recall that the dynamics generated by the Hamiltonian in Eq.~\eqref{ham} results in an equilibrium state of the form~\cite{schwarzhans2025quantum}:
\begin{align*}    \rho_{\text{eq}}=\sum_{i=1}^{d_{S}} p_i\ket{i}\bra{i}_{S} \bigotimes_{l=1}^{N_{E}} \tilde{\rho}_l^{(i)},
\end{align*}
with the conditional states $\tilde{\rho}_{l}^{(i)}$ given by
\begin{align} \label{eCondStates}
    \tilde{\rho}_{l}^{(i)} = \sum_{n_{l}} \bra{E_{n_{l}}^{(i)}}\tilde{\rho}_{l,0}\ket{E_{n_{l}}^{(i)}} \ket{E_{n_{l}}^{(i)}}\bra{E_{n_{l}}^{(i)}},
\end{align}
where $H_{l}^{(i)}=\sum_{n_{l}} E_{n_{l}}^{(i)} \ket{E_{n_{l}}^{(i)}}\bra{E_{n_{l}}^{(i)}}$ is the conditional Hamiltonian corresponding to the measurement outcome $i$. We now prove that
\begin{align} \label{eUnitInvar}
    \Big\langle F\Big(\mathbb{P}_{H^{(i)}} [\rho_{0}], \mathbb{P}_{H^{(j)}} [\rho_{0}]\Big) \Big\rangle_\text{GUE} = \Big\langle F\Big(\mathbb{P}_{H^{(i)}} [U\rho_{0}U^{\dag}], \mathbb{P}_{H^{(j)}} [U\rho_{0}U^{\dag}]\Big) \Big\rangle_\text{GUE}.
\end{align}
where $U$ is an arbitrary unitary transform, and the conditional state $\tilde{\rho}_l^{\left(i\right)}$ is written in terms of the pinching map with respect to $H^{(i)}$, such that $\rho^{(i)}_l = \mathbb{P}_{H^{(i)}} \left[ \rho_{0} \right]=\sum_{n_{l}}  \ket{E_{n_{l}}^{(i)}}\bra{E_{n_{l}}^{(i)}} \rho_0 \ket{E_{n_{l}}^{(i)}}\bra{E_{n_{l}}^{(i)}}$.

As a first step, recall the rotational invariance property of the Gaussian Unitary Ensemble (GUE)~\cite{Livan2018}, namely that a Hamiltonian $H^{(i)}$ is selected from the ensemble with the same probability as $U^{\dag}H^{(i)}U$ for any unitary operator $U$. In our notation, we can write this as
\begin{align} \label{eUnitInv_eigs}
    \sum_{n} E_{n}^{(i)} \ket{E_{n}^{(i)}}\bra{E_{n}^{(i)}} \preq \sum_{n} E_{n}^{(i)} U^{\dag}\ket{E_{n}^{(i)}}\bra{E_{n}^{(i)}}U \equiv \sum_{n} E_{n}^{(i)} \ket{\tilde{E}_{n}^{(i)}}\bra{\tilde{E}_{n}^{(i)}}, 
\end{align}
where $\preq$ denotes that the two operators are associated with the same probability, and $\ket{\tilde{E}_{n}^{(i)}}:=U^{\dag}\ket{E_{n}^{(i)}}$. Combining this with Eq.~\eqref{eCondStates}, we see that
\begin{align*}
    \mathbb{P}_{H^{(i)}} \left[ \rho_{0} \right] \preq \sum_{n} \bra{\tilde{E}_{n}^{(i)}}\rho_{0}\ket{\tilde{E}_{n}^{(i)}} \ket{\tilde{E}_{n}^{(i)}}\bra{\tilde{E}_{n}^{(i)}} = U^{\dag} \left( \sum_{n} \bra{E_{n}^{(i)}}U\rho_{0}U^{\dag}\ket{E_{n}^{(i)}} \ket{E_{n}^{(i)}}\bra{E_{n}^{(i)}} \right) U = U^{\dag} \mathbb{P}_{H^{(i)}} \left[ U\rho_{0}U^{\dag}\right] U ,
\end{align*}
and therefore that
\begin{align*}
    F\Big(\mathbb{P}_{H^{(i)}} [\rho_{0}], \mathbb{P}_{H^{(j)}} [\rho_{0}]\Big) \preq F\Big(U^{\dag} \mathbb{P}_{H^{(i)}} \left[ U\rho_{0}U^{\dag}\right] U, U^{\dag} \mathbb{P}_{H^{(j)}} \left[ U\rho_{0}U^{\dag}\right] U\Big) .
\end{align*}
Noting that the fidelity is invariant under a unitary transformation of both arguments, we find that Eq.~\eqref{eUnitInvar} holds, concluding the proof.

\subsection{Finite fidelity limit for pure initial states} \label{sFiniteFidLimit}

Here we give a heuristic argument as to why the average fidelity between conditional states does not vanish with increasing environment dimension in the case of a pure state. First let us use Eq.~\eqref{eCondStates} to write out the fidelity explicitly in this case (again, omitting the tilde and subscript $l$ for clarity):
\begin{align} \label{eCondFid}
    F\left(\rho^{(i)}, \rho^{(j)}\right) = \operatorname{tr} \left(  \sqrt{\sum_{n,n',m} \sqrt{\bra{E_{n}^{(i)}}\rho_{0}\ket{E_{n}^{(i)}}} \sqrt{\bra{E_{n'}^{(i)}}\rho_{0}\ket{E_{n'}^{(i)}}} \langle E_{m}^{(j)} |\rho_{0}| E_{m}^{(j)} \rangle \langle E_{n}^{(i)} | E_{m}^{(j)} \rangle \langle E_{m}^{(j)} | E_{n'}^{(i)} \rangle \ket{E_{n}^{(i)}}\bra{E_{n'}^{(i)}}}\right)^{2} .
\end{align}
Now, since $\ket{E_{n}^{(i)}}$ is an eigenvector of $H^{(i)}_{l}$, a matrix randomly selected according to the GUE, its elements in an arbitrary basis $\lbrace \ket{\alpha}\in \mathcal{H}_{l} \rbrace_{\alpha}$ can be treated as i.i.d.\ random variables in the large-$d_{l}$ limit. In particular, defining the variable $y:=d_{l} \vert \langle E_{n}^{i} | \alpha \rangle \vert^{2}$, they are distributed according to the probability density $p(y)=e^{-y}$~\cite{truong2016statistics}. Let us now examine the elements in the sum in Eq.~\eqref{eCondFid} individually in the large-$d_{l}$ limit. 

First, we know from Appendix~\ref{sUnitInvariance} that the average fidelity is invariant under unitary rotations of the initial state, and therefore for $\rho_{0}$ pure, the average does not depend on the specific pure state chosen. Consequently, we may write $\rho_{0}=\ket{\alpha}\bra{\alpha}$ for arbitrary $\ket{\alpha}$, and hence treat $\bra{E_{n}^{(i)}}\rho_{0}\ket{E_{n}^{(i)}}$ as the variable $\vert \langle E_{n}^{(i)} | \alpha \rangle \vert^{2}=y/d_{l}$. Likewise, we can treat $|\langle E_{n}^{(i)} | E_{m}^{(j)} \rangle|$ as the variable $\vert \langle E_{n}^{(i)} | \alpha \rangle \vert=\sqrt{y/d_{l}}$. 

Noting that under permutation of the appropriate labels, this accounts for all factors in the sum in Eq.~\eqref{eCondFid}, we can invoke the i.i.d.\ assumption to get a crude estimate of the average fidelity in the large-$d_{l}$ limit, replacing each factor with its average magnitude to get
\begin{align*}
    \left\langle  F\left(\rho^{(i)}, \rho^{(j)}\right) \right\rangle_\mathrm{GUE} &\sim \operatorname{tr} \left(  \sqrt{\sum_{n,n',m} \left\langle\sqrt{\frac{y}{d_{l}}}\right\rangle_\mathrm{GUE} \left\langle\sqrt{\frac{y}{d_{l}}}\right\rangle_\mathrm{GUE} \left\langle\frac{y}{d_{l}}\right\rangle_\mathrm{GUE} \left\langle\sqrt{\frac{y}{d_{l}}}\right\rangle_\mathrm{GUE} \left\langle\sqrt{\frac{y}{d_{l}}}\right\rangle_\mathrm{GUE} \ket{E_{n}^{(i)}}\bra{E_{n'}^{(i)}}}\right)^{2} \\
    &= d_{l} \left\langle\sqrt{\frac{y}{d_{l}}}\right\rangle^{4}_\mathrm{GUE}  \left\langle\frac{y}{d_{l}}\right\rangle_\mathrm{GUE} \, \operatorname{tr} \left(  \sqrt{\sum_{n,n'}  \ket{E_{n}^{(i)}}\bra{E_{n'}^{(i)}}}\right)^{2} \\
    &= \frac{1}{d_{l}^{2}} \left\langle\sqrt{y}\right\rangle^{4}_\mathrm{GUE}  \left\langle y \right\rangle_\mathrm{GUE} \operatorname{tr} \left( \sum_{n,n'}  \ket{E_{n}^{(i)}}\bra{E_{n'}^{(i)}}\right)^{2} \\
    &= \left\langle\sqrt{y}\right\rangle^{4}_\mathrm{GUE}  \left\langle y \right\rangle_\mathrm{GUE}.
\end{align*}
Using the probability density $p(y)=e^{-y}$ to calculate the averages in the above expression, one finds the limit
\begin{align*}
    \left\langle  F\left(\rho^{(i)}, \rho^{(j)}\right) \right\rangle_\mathrm{GUE} &\sim \lim_{d_{l}\to\infty} \left( \left\langle\sqrt{y}\right\rangle^{4}_\mathrm{GUE}  \left\langle y \right\rangle_\mathrm{GUE} \right) = \frac{\pi^{2}}{16} \approx 0.62 ,
\end{align*}
which, despite the crudeness of the approximation used, agrees very well with the result shown in Fig.~\ref{fig:fidelities}a (as $\left\langle \mathcal{E}_{\text{obj}} \right \rangle_{\text{GUE}} = \left\langle F\left(\rho^{(i)},\rho^{(j)}\right)\right\rangle_{\text{GUE}}$ for our chosen initial states).

\section{Properties of the average effective dimension with respect to the Gaussian unitary ensemble}

\subsection{Unitary invariance of the effective dimension with respect to initial state} \label{sUnit_invar_eff_dim}
Here we show that the term $\left\langle \mathcal{E}_{\text{eq}} \right \rangle_{\text{GUE}}$ in the error bound (Eq.~\eqref{error_bound_GUE}) is invariant under a unitary transformation on the initial state of a sub-environment $l$, i.e.
\begin{align} \label{eq:unit_invar_deff}
    \left\langle\frac{1}{ \sqrt{d_{\mathrm{eff}}(\rho_{l,0})}}\right\rangle_{\text{GUE}} &= \left\langle\frac{1}{ \sqrt{d_{\mathrm{eff}}(U \rho_{l,0} U^\dag)}}\right\rangle_{\text{GUE}}
\end{align}
where $U$ is an arbitrary unitary transform and
\begin{align}
    \left\langle\frac{1}{ \sqrt{d_{\mathrm{eff}}(\rho_{l,0})}}\right\rangle_{\text{GUE}} = \left\langle\frac{1}{2}\sqrt{ \sum_{i}^{d_{S}} \prod_{l}^{N_E}\sum_{n_l}^{d_{l}} \bra{E_{n_l}^{(i)}} \rho_{l,0} \ket{E_{n_l}^{(i)}}^{2} }\right\rangle_{\text{GUE}}.
\end{align}
From Eq.~\eqref{eUnitInv_eigs}, we know that 
\begin{align*}
    \sum_{n} E_{n}^{(i)} \ket{E_{n}^{(i)}}\bra{E_{n}^{(i)}} \preq \sum_{n} E_{n}^{(i)} U^{\dag}\ket{E_{n}^{(i)}}\bra{E_{n}^{(i)}}U
\end{align*}
which implies
\begin{align*}
    \bra{E_{n_l}^{(i)}} \rho_{l,0} \ket{E_{n_l}^{(i)}}^{2} \preq \bra{E_{n_l}^{(i)}} U \rho_{l,0} U^{\dag} \ket{E_{n_l}^{(i)}}^{2} = \bra{E_{n_l}^{(i)}} \tilde{\rho}_{l,0} \ket{E_{n_l}^{(i)}}^{2}.
\end{align*}
The unitary invariance of $\left\langle \mathcal{E}_{\text{obj}} \right \rangle_{\text{GUE}}$, given by Eq.~\eqref{eq:unit_invar_deff} follows directly.

\subsection{Maximised effective dimension for maximally mixed initial sub-environment states} \label{SM:max_mixed_initial_state}
We now calculate the effective dimension for a system-environment whose sub-environments are initially maximally-mixed:
\begin{align*}
    \rho(0)= \ket{\psi_{S,0}}\!\bra{\psi_{S,0}}\bigotimes_{l=1}^{N_E}  \frac{1}{d_{l}} \mathbb{1}_{l},
\end{align*}
and we assume the system $S$ is initially in an equal superposition in the measurement basis $|\psi_{S,0}\rangle = \frac{1}{\sqrt{d_S}} \sum_{i=1}^{d_S} |i\rangle$
Using Eq.~\eqref{eq:simplified_deff}, we can see that this choice of initial state maximises the effective dimension:
\begin{align*}
    d_{\mathrm{eff}} & = \left[\sum_{i=1}^{d_{S}}p_{i}^{2} \prod_{l=1}^{N_E}\sum_{n_l =1}^{d_{l}} \left\lbrace\bra{E_{n_l}^{(i)}} \left(  \frac{1}{d_{l}}\mathbb{1}_{l} \right)  \ket{E_{n_l}^{(i)}}\right\rbrace^{2}    \right]^{-1}\\
    & =\left[\sum_{i=1}^{d_{S}} p_{i}^{2} \prod_{l=1}^{N_E}\sum_{n_l =1}^{d_{l}}  \left(\frac{1}{d_{l}}\right)^{2} \right]^{-1} \\
    & = \left[\sum_{i=1}^{d_{S}} p_{i}^{2} \prod_{l=1}^{N_E}  \left(\frac{1}{d_{l}}\right)  \right]^{-1} \\
    & = \left[\sum_{i=1}^{d_{S}} p_{i}^{2}  \left(\frac{1}{d_{l}}\right)^{N_E}  \right]^{-1} \\
    & = \left[ \left(\frac{1}{d_{S}}\right) \left(\frac{1}{d_{l}}\right)^{N_E}  \right]^{-1} \\
    & = d_{S} \cdot (d_{l})^{N_E},
\end{align*}
where we use that $p_{i}=\frac{1}{d_{S}}$ and $N_E$ is the total number of sub-environments. It is also worth noting that the equilibrium state of each sub-environment is also the maximally mixed state, meaning $\rho_{l,0}=\rho_{l}^{(i)}=\rho_{l}^{(j)}$. Therefore, observers cannot distinguish between system outcomes, no matter how they group together sub-environments into observer systems. There is no objectivity and the equilibrium state is very far from an SBS state.

\subsection{Effect on fidelity and effective dimension under choice of thermal Hamiltonian for thermal initial states} \label{SM:thermal}

In this section, we outline the effect of the thermal state Hamiltonian on both terms $\left\langle \mathcal{E}_{\text{obj}} \right \rangle_{\text{GUE}}$ and $\left\langle \mathcal{E}_{\text{eq}} \right \rangle_{\text{GUE}}$ in the error bound. Our initial state is 
\begin{align*}
    \rho(0) = \rho_{S,0} \bigotimes_{l=1}^{N_E} \rho_{l,0},
\end{align*}
where the system is a qubit in an equal superposition $\rho_{S,0} = \ket{\psi_{S,0}}\bra{\psi_{S,0}}$, such that $\ket{\psi_{S,0}} = \frac{1}{\sqrt{2}}\ket{0}+\ket{1}$ and each environment is thermal $\rho_{l,0} = \rho_{th}$. We begin by defining a thermal state as $\rho_{th} = \frac{e^{-\beta H_{th}}}{Z}$, with respect to some Hamiltonian $H_{th}$, where $Z = \operatorname{tr}\left( e^{-\beta H_{th}} \right)$. Noting that, under the transformation $H_{th} \to UH_{th}U^{\dag}$ where $U$ is an abitrary unitary operator, we have $e^{-\beta H_{th}}\to Ue^{-\beta H_{th}}U^{\dag}$, and recalling the unitary invariance of the trace operation, we see that $\rho_{th}\to U\rho_{th}U^{\dag}$ under $H_{th} \to UH_{th}U^{\dag}$. We showed in Appendix \ref{sUnit_invar_eff_dim} (\ref{sUnitInvariance}) the invariance of $\left\langle \mathcal{E}_{\text{eq}} \right \rangle_{\text{GUE}}$ ($\left\langle \mathcal{E}_{\text{obj}} \right \rangle_{\text{GUE}}$) with respect to unitary transformations of the initial observer system state. Any Hermitian matrix with the same spectrum can be related by a unitary transformation. 

Therefore, we can conclude that the choice of thermal Hamiltonian eigenstates has no effect, in our numerical simulations, on the error bound when averaged over the GUE. If we consider thermal Hamiltonians with different eigenenergy spectrums, they cannot be related by a unitary transformation and so this will affect our results. The only effect is on the temperature scaling. In the following Appendix, we address this and explain our choice of temperature in Fig.~\ref{fig:thermal_states}.

\subsection{Details of thermal states used in numerical simulations} \label{SM:thermal_num}

As outlined in the previous Appendix (\ref{SM:thermal}), only the distribution of eigenenergies of the thermal Hamiltonian defining the sub-environment initial states will affect the error bound, when averaged over the GUE. For that reason, we choose to model an optical thermal field, due to the simplicity of the thermal Hamiltonian $H_{th} = \hbar \omega \left(\hat{a}^{\dag} \hat{a} + \frac{1}{2}  \right)$, with eigenenergies $E_n = \hbar\omega \left(n+\frac{1}{2}\right)$, such that all energy gaps are fixed to $\hbar\omega$. The density operator for the thermal field is \cite{gerry2023introductory}:
\begin{equation}
    \rho_{th} = \frac{e^{-H_{th}\beta}}{Z} = \sum_{n=0}^{\infty} P_{n} \ket{n}\bra{n},
\end{equation}
where $\beta=1/K_{B}T$ is the inverse temperature and the probability that a mode is excited to the $n$th level is $P_{n} = \frac{1}{Z} e^{-E_{n}\beta}$. This results in an average photon number of
\begin{equation}
    \bar{n} = \frac{1}{e^{\hbar\omega \beta}-1}.
\end{equation}
From this, we can write
\begin{equation}
    \beta \omega = \frac{1}{\hbar} \operatorname{ln}\left( \frac{1+\bar{n}}{\bar{n}} \right)
\end{equation}
and define the thermal state in terms of $\bar{n}$ 
\begin{equation} \label{eq:textbook_th}
    \rho_{th} = \sum_{n=0}^{\infty} \frac{\bar{n}^{n}}{(1+\bar{n})^{n+1}} \ket{n}\bra{n}
\end{equation}

To perform our numerical simulations, we need to write the thermal state as a finite density matrix. To do this, we use a built-in QuTiP function \cite{johansson2012qutip} which defines the thermal state with $d$ dimensions as
\begin{equation}
    \rho_{th} = \frac{1}{\operatorname{tr}(\rho_{th})}   \sum_{n=0}^{d-1} \left(\frac{\bar{n}}{1+\bar{n}}\right)^{n} \ket{n}\bra{n},
\end{equation}
where $\operatorname{tr}(\rho_{th})$ indicates the renormalisation of the density matrix for finite dimension $d$. In the simulations, the parameter we define is the average photon number $\bar{n}$, so by varying this we are varying the dimensionless quantity $\beta \omega$, without ever defining $\beta$ or $\omega$ individually.

\end{document}